\documentclass[12pt,eqno]{article}
\usepackage{amsmath,amssymb}
\numberwithin{equation}{section}

\newcommand{\alp}{\alpha}
\newcommand{\bt}{\beta}

\newcommand{\dlt}{\delta}

\newcommand{\vep}{\varepsilon}
\newcommand{\tht}{\theta}
\newcommand{\btht}{\bar{\tht}}

\newcommand{\lmd}{\lambda}
\newcommand{\Lmd}{\Lambda}
\newcommand{\sgm}{\sigma}
\newcommand{\Sgm}{\Sigma}
\newcommand{\vph}{\varphi}

\newcommand{\dalp}{\dot{\alpha}}

\newcommand{\vth}{\vartheta^*}

\newcommand{\be}{\begin{equation}}
\newcommand{\ee}{\end{equation}}
\newcommand{\bea}{\begin{eqnarray}}
\newcommand{\eea}{\end{eqnarray}}
\newcommand{\eql}{\!\!\!&=\!\!\!&}

\newcommand{\defa}{\!\!\!&\equiv\!\!\!&}

\newcommand{\mtrx}[4]{\brkt{\begin{array}{cc}#1&#2\\#3&#4\end{array}}}
\newcommand{\dgnl}[2]{\brkt{\begin{array}{cc}#1& \\ &#2\end{array}}}
\newcommand{\vct}[2]{\brkt{\begin{array}{c}#1\\#2\end{array}}}

\newcommand{\tl}[1]{\tilde{#1}}
\newcommand{\bdm}[1]{{\mbox{\boldmath $#1$}}}

\newcommand{\der}{\partial}
\newcommand{\dr}{\!\!d}
\newcommand{\hc}{{\rm h.c.}}
\newcommand{\ie}{{\it i.e.}}

\newcommand{\vev}[1]{\langle #1 \rangle}

\newcommand{\brkt}[1]{\left( #1 \right)}
\newcommand{\brc}[1]{\left\{ #1 \right\}}
\newcommand{\sbk}[1]{\left[ #1 \right]}
\newcommand{\abs}[1]{\left| #1 \right|}

\renewcommand{\Re}{{\rm Re}}
\renewcommand{\Im}{{\rm Im}}

\newcommand{\cF}{{\cal F}}
\newcommand{\cG}{{\cal G}}
\newcommand{\cH}{{\cal H}}

\newcommand{\cL}{{\cal L}}

\newcommand{\cN}{{\cal N}}
\newcommand{\cO}{{\cal O}}
\newcommand{\cP}{{\cal P}}
\newcommand{\cQ}{{\cal Q}}

\newcommand{\cT}{{\cal T}}
\newcommand{\cV}{{\cal V}}
\newcommand{\cW}{{\cal W}}

\renewcommand{\ge}[2]{e_{#1}^{\;\;#2}}
\newcommand{\udl}[1]{\underline{#1}}
\newcommand{\nV}{n_{\rm V}}
\newcommand{\nVp}{n'_{\rm V}}
\newcommand{\nH}{n_{\rm H}}
\newcommand{\nC}{n_{\rm C}}
\newcommand{\dmx}{d_{a}^{\;\;b}}

\newcommand{\SUu}{SU(2)_{\mbox{\scriptsize $\bdm{U}$}}}

\newcommand{\bPhi}{\bdm{\Phi}}
\newcommand{\bvph}{\bdm{\vph}}
\newcommand{\Ip}{I'_{\rm nq}}
\newcommand{\gc}{g^{\rm c}}
\newcommand{\gh}{g^{\rm h}}

\newcommand{\NP}[1]{{\it Nucl.~Phys.}~{\bf #1}}
\newcommand{\PL}[1]{{\it Phys.~Lett.}~{\bf #1}}

\newcommand{\PR}[1]{{\it Phys.~Rev.}~{\bf #1}}
\newcommand{\PRL}[1]{{\it Phys.~Rev.~Lett.}~{\bf #1}}
\newcommand{\PTP}[1]{{\it Prog.~Theor.~Phys.}~{\bf #1}}

\newcommand{\JH}[1]{{\it JHEP}~{\bf #1}}

\begin{document}

\begin{titlepage}
\null
\begin{flushright}
 {\tt hep-th/0610234}\\
YITP-06-53
\\
OU-HET 568/2006
\\
September, 2011
\end{flushright}

\vskip 2cm
\begin{center}
\baselineskip 0.8cm
{\LARGE \bf Roles of $Z_2$-odd $N=1$ multiplets 
in off-shell dimensional reduction of 5D supergravity
}

\lineskip .75em
\vskip 2.5cm

\normalsize

{\large\bf Hiroyuki Abe}${}^1\!${\def\thefootnote{\fnsymbol{footnote}}
\footnote[1]{\it e-mail address:abe@yukawa.kyoto-u.ac.jp}}
{\large\bf and Yutaka Sakamura}${}^2\!${\def\thefootnote{\fnsymbol{footnote}}
\footnote[2]{\it e-mail address:sakamura@het.phys.sci.osaka-u.ac.jp}}

\vskip 1.5em

${}^1${\it Yukawa institute for theoretical physics, Kyoto University, \\ 
Kyoto 606-8502, Japan}

\vskip 1.0em

${}^2${\it Department of Physics, Osaka University, \\ 
Toyonaka, Osaka 560-0043, Japan}

\vspace{18mm}

{\bf Abstract}\\[5mm]
{\parbox{13cm}{\hspace{5mm} \small
We discuss the dimensional reduction of five-dimensional supergravity 
compactified on $S^1/Z_2$ keeping the $N=1$ off-shell structure. 
Especially we clarify the roles of the $Z_2$-odd $N=1$ multiplets in  
such an {\it off-shell dimensional reduction}. 
Their equations of motion provide constraints on the $Z_2$-even 
multiplets and extract the zero modes from the latter. 
The procedure can be applied to wide range of models 
and performed {\it in a background-independent way}. 
We demonstrate it in some specific models. 
}}

\end{center}

\end{titlepage}

\clearpage

\section{Introduction}
Higher dimensional supergravity (SUGRA) has been attracted much attention 
and extensively investigated in various aspects, such as 
effective theories of the superstring theory or M-theory~\cite{polchinski}, 
AdS/CFT correspondence~\cite{maldacena}, 
the model building in the context of the brane-world scenario 
(see \cite{braneworld}, for example), etc.  
Among them, five-dimensional (5D) supergravity~\footnote{
We focus on 5D SUGRA with eight supercharges in this paper. }
compactified on an orbifold~$S^1/Z_2$ has been thoroughly investigated 
since it is shown to appear as an effective theory of the strongly coupled 
heterotic string theory~\cite{HW} compactified 
on a Calabi-Yau 3-fold~\cite{LOSW}. 
Besides, the supersymmetric (SUSY) extensions of 
the Randall-Sundrum model~\cite{RS} 
are also constructed in 5D SUGRA on $S^1/Z_2$~\cite{GP,FLP,ABN}. 
Their low-energy effective theories become 
four-dimensional (4D) supergravity as the fifth dimension is compactified. 
In this paper, we will discuss the dimensional reduction of 
5D SUGRA on $S^1/Z_2$ to derive its 4D effective theory. 

The conformal SUGRA formulation~\cite{KU,zucker,KO2,FKO,KO3}, 
which is one of the off-shell formulation of SUGRA, 
is quite useful to construct the 5D SUGRA action, 
especially in a case that 
there are terms localized on the orbifold boundaries. 
The formulation is systematic and straightforward. 
Firstly we construct a 5D superconformal invariant action. 
Then we fix extra superconformal symmetries by imposing the gauge-fixing 
conditions so that we obtain the desired Poincar\'{e} SUGRA action.  
The conformal SUGRA formulation is also useful 
to describe the 4D effective theory of 5D SUGRA 
because it makes $N=1$ SUSY~\footnote{
$N=1$ denotes supersymmetry with four supercharges in this paper. } 
preserved by the orbifold projection manifest and also enables to discuss 
the mediation of SUSY breaking effects in a transparent way 
when SUSY is spontaneously broken. 
Such an off-shell 4D effective action is first derived in Ref.~\cite{LS}. 
Although it is useful and instructive, it covers only a specific case 
where the background geometry is the Randall-Sundrum warped spacetime. 
Besides it is not clear how the off-shell 4D action is related 
to the original off-shell 5D SUGRA action in the derivation of Ref.~\cite{LS}. 
In order to clarify their relation, 
we need to derive the former directly from 
the latter {\it keeping the $N=1$ off-shell structure}. 
For this purpose, it is convenient to decompose each 5D superconformal 
multiplet into $N=1$ multiplets~\cite{KO3}. 
Then we can rewrite the 5D conformal SUGRA action in terms of 
$N=1$ multiplets in the language of $N=1$ off-shell SUGRA~\cite{PST1,AS1}. 
The formulation of such an $N=1$ off-shell action is pioneered by 
the authors of Ref.~\cite{LLP} for the minimal 5D SUGRA at the linearized level. 
Their formulation is useful for the calculation of the SUGRA loop 
contributions to the visible sector 
in the 5D braneworld models~\cite{BGGLLNP}. 
Our purpose in this paper is to derive the off-shell 4D action 
{\it at the full SUGRA level}. 
As we mentioned in Ref.~\cite{AS3}, however, it is a nontrivial task 
to derive it because there are some multiplets, 
such as the compensator multiplet, that cannot be naively expanded 
into the Kaluza-Klein (K.K.) modes keeping the $N=1$ structure. 

Recently the authors of Ref.~\cite{PST2} proposed 
a systematic method to derive the effective action of 5D SUGRA 
as an off-shell 4D SUGRA action avoiding the above difficulty. 
Their method is quite useful since 
it is applicable to wide range of 5D SUGRA models. 
In this paper, we will reinterpret their method 
to clarify the guiding principle of the procedure, and modify it. 
Our modification makes it possible to apply this method 
to more general class of models, and clarifies the limits of validity for it. 
Especially we modify a treatment of $N=1$ multiplets which are odd 
under the $Z_2$-parity of the orbifold. 
These multiplets are simply dropped by hand in Ref.~\cite{PST2} 
because they do not contain light 4D modes which 
appear in the low-energy effective theory. 
In general, however, such heavy fields affect the effective theory 
when they are integrated out~\cite{LS,HLW}. 
We treat the $Z_2$-odd multiplets carefully and 
show that their equations of motion provide constraints on 
the $Z_2$-even multiplets and extract the zero mode from the latter. 
We show that the procedure of the off-shell dimensional reduction 
can be performed {\it in a background-independent way}. 
This is no surprise 
because the information on the background is obtained 
only after moving to the on-shell description while the $N=1$ off-shell 
structure is kept during the procedure. 
We demonstrate this off-shell dimensional reduction explicitly 
in some specific models. 

The paper is organized as follows. 
We review the $N=1$ description of the 5D conformal SUGRA action and 
provide the boundary actions in a case that the compensator consists of 
one or two hypermultiplets. 
In Sect.~\ref{ODR}, we explain the procedure of the off-shell dimensional 
reduction in detail. 
In Sect.~\ref{sp_ex}, we derive 4D effective theories of some specific models 
by applying the procedure provided in Sect.~\ref{ODR}. 
Sect.~\ref{comments} is devoted to the summary and some comments. 
In Appendix~\ref{SCgf}, the gauge fixing conditions for 
the extra superconformal symmetries are listed in our notation. 
In Appendix~\ref{3kI_neq_2mI}, we provide a complement 
of Sect.~\ref{VI_integrate}.

\section{$N=1$ off-shell description of 5D SUGRA action}
\subsection{$N=1$ multiplets and bulk action}
Throughout this paper, we use $\mu,\nu,\cdots=0,1,2,3$ for the 4D world 
vector indices and $y$ for the coordinate of 
the fifth dimension compactified on $S^1/Z_2$. 
The corresponding local Lorentz indices are denoted by underbarred indices. 
We take the fundamental region of the orbifold as $0\leq y\leq \pi R$, 
where $R$ is a constant,\footnote{
In principle, $R$ is nothing to do with the radius of the orbifold~$r$, 
but it is convenient to take it so that it coincides with the latter 
after the radius is stabilized. }
and take the unit of $M_5=1$, where $M_5$ is the 5D Planck mass. 

We assume the following form of the metric, 
\be
 ds_5^2 = e^{2\sgm(y)}g_{\mu\nu}dx^\mu dx^\nu-\brkt{\ge{y}{4}dy}^2, 
\ee
where $\sgm(y)$ is a warp factor, which is a function of only $y$. 
The off-diagonal components of the metric~$g_{\mu y}$ can always be 
gauged away. 

According to Ref.~\cite{KO3}, each 5D superconformal multiplet can be 
decomposed into $N=1$ superconformal multiplets as follows. 
The 5D Weyl multiplet is decomposed into the $N=1$ Weyl 
multiplet~$E_{\rm W}=(\ge{\mu}{\udl{\nu}},\psi_{\mu+}^\alp,\cdots)$ and 
a real general multiplet~\footnote{In the notation of Ref.~\cite{KO3}, 
this general multiplet is denoted as $\bdm{W}_y$. } 
\be
 V_E = (\ge{y}{4},-2\psi_{y-},-2V_y^2,2V_y^1,\cdots), \label{V_E}
\ee
where $V_M^{t=1,2,3}$ ($M=\mu,y$) are the $\SUu$ gauge fields, 
which are auxiliary fields. 
As mentioned in Ref.~\cite{KO3}, 
the off-diagonal components of the Weyl multiplet, which are $Z_2$-odd, 
do not form an $N=1$ multiplet 
but appear in the covariant derivatives of the $Z_2$-even components 
in an unusual way. 
We will not discuss them any more since they are essentially irrelevant to 
the following discussion. 
Each 5D vector multiplet is decomposed into 4D vector and chiral 
multiplets~$V^I$ and $\Sgm^I$ ($I=0,1,\cdots,\nV$),\footnote{
$\Sgm^I$ is related to $\Phi_S^I$ in our previous works~\cite{AS1,AS3,AS2} 
as $\Sgm^I=-i\Phi_S^I$. } 
and each hypermultiplet is decomposed into two chiral 
multiplets~$(\Phi^{2\hat{a}-1},\Phi^{2\hat{a}})$ 
($\hat{a}=1,\cdots,\nC+\nH$). 
Here $\nC$ and $\nH$ are numbers of 
the compensator and the physical hypermultiplets, respectively. 
The Weyl weight of each multiplet is listed in Table~\ref{Weyl_weight}. 
\begin{table}[t]
\begin{center}
\begin{tabular}{|c|c|c|c|c|c|} \hline
\rule[-2mm]{0mm}{7mm} & $E_{\rm W}$ & $V_E$ & $V^I$ & $\Sgm^I$ & $\Phi^a$  
\\ \hline 
Weyl weight & $-1$ & $-1$ & $0$ & $0$ & $3/2$ \\ \hline
\end{tabular}
\end{center}
\caption{The Weyl weight of each multiplet. 
The indices run over $I=0,1,\cdots,\nV$ and $a=1,2,\cdots,2(\nC+\nH)$. }
\label{Weyl_weight}
\end{table}
The value in the table denotes the Weyl weight of the lowest component 
in each multiplet. 
The Weyl weight of higher component rises by $1/2$. 
In other words, the Grassmann variable~$\tht^\alp$ has the Weyl weight $-1/2$ 
in the superfield language. 

Using these $N=1$ multiplets, the 5D conformal SUGRA action can be 
expressed as the following $N=1$ superspace action~\cite{PST1,AS1}. 
\bea
 S \eql \int\dr^5x\;\brc{\cL_{\rm vector}+\cL_{\rm hyper}
 +\sum_{\vth=0,\pi}\cL^{(\vth)}_{\rm brane}\dlt(y-\vth R)}, \nonumber\\
 \cL_{\rm vector} \eql \sbk{\int\dr^2\tht\;\brc{-\frac{\cN_{JK}}{4}(\Sgm)
 \cW^J\cW^K+\frac{\cN_{IJK}}{48}\bar{D}^2
 \brkt{V^ID^\alp\der_y V^J-D^\alp V^I\der_y V^J}\cW^K_\alp}+\hc} \nonumber\\
 &&-e^{2\sgm}\int\dr^4\tht\;V_E^{-2}\cN(\cV), \nonumber\\
 \cL_{\rm hyper} \eql -2e^{2\sgm}\int\dr^4\tht\; V_E\dmx
 \bar{\Phi}^b\brkt{e^{-2igV^It_I}}^a_{\;\;c}\Phi^c \nonumber\\
 &&-e^{3\sgm}\sbk{\int\dr^2\tht\;\Phi^a\dmx\rho_{bc}
 \brkt{\der_y-2ig\Sgm^It_I}^c_{\;\;d}\Phi^d+\hc}, 
 \label{action1}
\eea
where $a,b,\cdots =1,2,\cdots,2(\nC+\nH)$. 
The {\it norm function}~$\cN$ is a cubic function defined by 
\be
 \cN(X) \equiv C_{IJK}X^IX^JX^K, \label{def_normf}
\ee
where $C_{IJK}$ is a real constant tensor which is completely 
symmetric for the indices, and  
$\cN_I\equiv\der\cN/\der X^I$, $\cN_{IJ}\equiv\der^2\cN/\der X^I\der X^J$, 
$\cdots$. 
The metric of the hyperscalar space~$\dmx$ can be brought 
into the standard form~\cite{WLP}
\be
 \dmx = \dgnl{\bdm{1}_{2\nC}}{-\bdm{1}_{2\nH}}, 
\ee
and an antisymmetric tensor~$\rho_{ab}$ is defined as 
$\rho_{ab}\equiv i\sgm_2\otimes\bdm{1}_{\nC+\nH}$. 
The superfields~$\cW_\alp^I$ and $\cV^I$ are defined by~\footnote{
The gauge invariant combinations for $\SUu$ 
are $\cW_\alp^I$ and $\cV^I/V_E$. } 
\bea
 \cW_\alp^I \defa -\frac{1}{4}\bar{D}^2D_\alp V^I, \nonumber\\
 \cV^I \defa -\der_y V^I+\Sgm^I+\bar{\Sgm}^I. 
\eea
For simplicity, we will consider only abelian gauge groups in this paper. 
The boundary Lagrangians~$\cL^{(\vth)}_{\rm brane}$ ($\vth=0,\pi$) are 
discussed in the next two subsections. 
The letter~$g$ symbolically denotes the gauge coupling constants. 
Their explicit forms are provided together with the generators~$t_I$, 
such as Eqs.(\ref{gt0}) and (\ref{nc2_igtI}). 
These gauge couplings can be either even or odd under the $Z_2$-parity. 
While $Z_2$-even couplings are constants over the whole spacetime, 
$Z_2$-odd couplings have kink profiles. 
Namely the latter are constants times 
the periodic step function~$\vep(y)$ defined by 
\be
 \vep(y) = \begin{cases} 1 & (2n\pi R < y < (2n+1)\pi R) \\
 0 & (y=n\pi R) \\ -1 & ((2n-1)\pi R < y < 2n\pi R) \end{cases} 
 \label{def_vep}
\ee
for an arbitrary integer~$n$. 
Such $Z_2$-odd coupling constants are consistently realized in SUGRA context 
by the so-called four-form mechanism proposed in Ref.~\cite{FKO,BKV}. 

Note that Eq.(\ref{action1}) is a shorthand expression for the full SUGRA 
action. 
We can always restore the full action by promoting $d^4\tht$ and $d^2\tht$ 
integrals to the $D$- and $F$-term formulae of 
$N=1$ conformal SUGRA formulation~\cite{KU}, 
which are compactly listed in Appendix~C of Ref.~\cite{KO3}. 

The superspace action~(\ref{action1}) has a similar form 
to that of Ref.~\cite{MP}. 
As we have pointed out in Ref.~\cite{AS1}, however, the latter action is  
not consistent with the 5D conformal SUGRA based 
on Ref.~\cite{KO2,FKO,KO3} 
especially in the case that a physical hyperscalar has a nontrivial 
background just like in Ref.~\cite{MO}. 

As we have indicated in Ref.~\cite{AS3}, there are some $N=1$ multiplets 
that interfere with a naive dimensional reduction keeping the $N=1$ 
off-shell structure. 
$V_E$ is one of them. 
The authors of Ref.~\cite{PST2} noticed that it can be easily integrated out 
because it does not have a kinetic term.\footnote{
This does not mean that $\ge{y}{4}$ is an auxiliary field. 
It is also contained in $\Sgm^I$ ($I=0,\cdots,\nV$), 
which have their own kinetic terms.
}
After integrating out $V_E$, the bulk part of the action~(\ref{action1}) 
is rewritten as 
\bea
 \cL_{\rm bulk} \eql \sbk{\int\dr^2\tht\;\brc{-\frac{\cN_{JK}}{4}(\Sgm)
 \cW^J\cW^K+\frac{\cN_{IJK}}{48}\bar{D}^2
 \brkt{V^ID^\alp\der_y V^J-D^\alp V^I\der_y V^J}\cW^K_\alp}+\hc} \nonumber\\
 &&-3e^{2\sgm}\int\dr^4\tht\;\cN^{1/3}(\cV)\brc{\dmx\bar{\Phi}^b
 \brkt{e^{-2igV^It_I}}^a_{\;\;c}\Phi^c}^{2/3} \nonumber\\
 &&-e^{3\sgm}\sbk{\int\dr^2\tht\;\Phi^a\dmx\rho_{bc}
 \brkt{\der_y-2ig\Sgm^It_I}^c_{\;\;d}\Phi^d+\hc}. 
 \label{action2}
\eea
Note that this procedure can be performed independently of the boundary actions 
because $V_E$ does not appear in them. 

The expression~(\ref{action2}) is useful for finding
a supersymmetric solution, or a BPS background. 
The Killing spinor equations, or the BPS equations, are expresssed 
as conditions that $D$- and $F$-components of all $N=1$ multiplets vanish 
\cite{PST1,AS1}. 
Thus we have only to pick up linear terms for the $D$- and $F$-components 
from the action and put them to zero in order to obtain the BPS equations. 
This greatly simplifies the calculation. 
Let us demonstrate this procedure in specific models 
in the cases of $\nC=1,2$.

\subsection{$\nC=1$ case} \label{nc1_case}
Let us first consider a case 
that the compensator multiplet consists of a single hypermultiplet. 
In the following, we divide the index~$I$ into $(I',I'')$ 
so that $V^{I'}$ and $V^{I''}$ are odd and even 
under the $Z_2$-parity respectively. 
The $Z_2$-parity of each $N=1$ multiplet is listed 
in Table~\ref{Z2_parity1}.\footnote{ 
For the hypermultiplets~$(\Phi^{2\hat{a}-1},\Phi^{2\hat{a}})$, 
we can always choose the $Z_2$-parities as listed in Table~\ref{Z2_parity1} 
by using $\SUu$. }
\begin{table}[t]
\begin{center}
\begin{tabular}{|c|c|c|c|c|c|c|} \hline
\rule[-2mm]{0mm}{7mm} & $V^{I'}$ & $\Sgm^{I'}$ & $V^{I''}$ & $\Sgm^{I''}$ &
$\Phi^{2\hat{a}-1}$ & $\Phi^{2\hat{a}}$  \\ \hline 
$Z_2$-parity & $-$ & $+$ & $+$ & $-$ & $-$ & $+$ \\ \hline
\end{tabular}
\end{center}
\caption{The orbifold parities in the case of $\nC=1$. 
The indices run over 
$I'=0,1,\cdots,\nVp$; $I''=\nVp+1,\nVp+2,\cdots,\nV$; 
$\hat{a}=1,2,\cdots,\nH+1$. }
\label{Z2_parity1}
\end{table}

The boundary Lagrangian in this case is written as 
\bea
 \cL^{(\vth)}_{\rm brane} \eql \sbk{\int\dr^2\tht\;\frac{1}{4}
 f^{(\vth)}_{\bar{I}\bar{J}}(Q)\cW^{\bar{I}}\cW^{\bar{J}}+\hc}
 -3e^{2\sgm}\int\dr^4\tht\;\abs{\Phi^{a=2}}^{4/3}
 \exp\brc{-K^{(\vth)}(\bar{Q},Q,U)/3} \nonumber\\
 &&+e^{3\sgm}\sbk{\int\dr^2\tht\;\brkt{\Phi^{a=2}}^2 P^{(\vth)}(Q)+\hc}, 
 \label{L_bd1}
\eea
where $U^{\bar{I}}$ and $Q^{\bar{a}}$ are 
$N=1$ vector and chiral multiplets. 
The barred indices~$\bar{I}$ and $\bar{a}$ run over 
not only the induced multiplets on the boundaries 
from the bulk but also boundary-localized multiplets, if any. 
$\cW^{\bar{I}}$ is a superfield strength of $U^{\bar{I}}$. 
Functions~$f^{(\vth)}_{\bar{I}\bar{J}}$, $K^{(\vth)}$ and $P^{(\vth)}$ are 
the gauge kinetic functions, the K\"{a}hler potentials and 
the superpotentials, respectively. 
Note that only $\Phi^{a=2}$ can appear as an $N=1$ chiral compensator 
multiplet in the boundary actions 
because $\Phi^{a=1}$ is $Z_2$-odd and vanishes on the boundaries. 
The powers of $\Phi^2$ are determined by the Weyl weight counting.\footnote{
The arguments of $D$- and $F$-term formulae must have the Weyl weight 
2 and 3, respectively. } 
Since the Weyl weights of the matter multiplets 
in the $N=1$ off-shell action must be zero~\cite{KU}, 
the bulk multiplets can appear in $\cL^{(\vth)}_{\rm brane}$ 
only in the form of 
\be
 U^I = V^I, \;\;\;\;\;
 Q^{\hat{a}} = \frac{\Phi^{2\hat{a}}}{\Phi^2}. \;\;\;\;\;
 (\hat{a}\geq 2)  \label{nc1_US}
\ee

As a specific example, let us consider a case that 
$(\nC,\nH)=(1,1)$ and the boundary actions are absent. 
The hypermultiplets are charged for $(V^I,\Sgm^I)$ as
\be
 igt_I=\sgm_3\otimes \vep(y)\dgnl{\frac{3}{2}k_I}{m_I}, \label{gt0}
\ee
where the Pauli matrix~$\sgm_3$ acts on 
each hypermultiplet~$(\Phi^{2\hat{a}-1},\Phi^{2\hat{a}})$, 
and $\vep(y)$ is defined in (\ref{def_vep}). 
We assume that all $V^I$ are $Z_2$-odd for simplicity (\ie, $I=I'$). 
The relevant part of the Lagrangian to obtain the BPS equations is
\bea
 \cL \eql -3e^{2\sgm}\int\dr^4\tht\;\cN^{1/3}(\cV)\brc{
 e^{-3k\cdot V}\abs{\Phi^1}^2+e^{3k\cdot V}\abs{\Phi^2}
 -e^{-2m\cdot V}\abs{\Phi^3}^2-e^{2m\cdot V}\abs{\Phi^4}^2}^{2/3} \nonumber\\
 &&-2e^{3\sgm}\sbk{\int\dr^2\tht\;\brc{\Phi^1\brkt{\der_y
 +\frac{3}{2}\dot{\sgm}+3k\cdot\Sgm}\Phi^2
 -\Phi^3\brkt{\der_y+\frac{3}{2}\dot{\sgm}+2m\cdot\Sgm}\Phi^4}+\hc} 
 \nonumber\\
 &&+\cdots
 \label{rel_L1}
\eea
where $\dot{\sgm}\equiv d\sgm/dy$ and $k\cdot V\equiv \vep(y)k_IV^I$, 
$m\cdot V\equiv \vep(y)m_IV^I$, $\cdots$. 
The BPS equations are read off as follows. 
\bea
 \frac{e^{-2\sgm}}{\cN^{1/3}}\der_y\brkt{e^{2\sgm}\frac{\cN_I}{\cN^{2/3}}}
 \eql 6\vep(y)k_I\brkt{\abs{\vph^1}^2-\abs{\vph^2}^2}
 -4\vep(y)m_I\brkt{\abs{\vph^3}^2-\abs{\vph^4}^2}, 
 \nonumber\\
 \vep(y)\brkt{3k_I\vph^1\vph^2-2m_I\vph^3\vph^4} \eql 0, \nonumber\\
 \brkt{\der_y+\frac{3}{2}\dot{\sgm}+3k\cdot\vph_\Sgm}\vph^2 \eql 0, 
 \;\;\;\;\;
 \brkt{\der_y+\frac{3}{2}\dot{\sgm}-3k\cdot\vph_\Sgm}\vph^1 = 0, 
 \nonumber\\
 \brkt{\der_y+\frac{3}{2}\dot{\sgm}+2m\cdot\vph_\Sgm}\vph^4 \eql 0, 
 \;\;\;\;\;
 \brkt{\der_y+\frac{3}{2}\dot{\sgm}-2m\cdot\vph_\Sgm}\vph^3
 = 0, \label{BPSeq1}
\eea
where $\vph^a$ ($a=1,2,3,4$) and $\vph_\Sgm^I$ are the lowest components of 
the chiral multiplets~$\Phi^a$ and $\Sgm^I$, respectively. 
The arguments of $\cN$'s in the first equation are $2\Re\,\vph_\Sgm$. 
The equations in (\ref{BPSeq1}) come from the conditions that 
the $D$-component of $V^I$, $F$-components of $\Sgm^I$, $\Phi^1$, $\Phi^2$, 
$\Phi^3$ and $\Phi^4$ vanish. 
By solving (\ref{BPSeq1}) partially with the superconformal 
gauge-fixing conditions in Appendix~\ref{SCgf}, 
we obtain the following equations for $0<y<\pi R$. 
\be
 e^{-2\sgm}\der_y\brkt{e^{2\sgm}\cN_I(2\Re\,\vph_\Sgm)}+6k_I
 +\frac{2(3k_I-2m_I)\abs{C_0}^2
 \exp\brc{2(3k-2m)\cdot\int_0^y\dr y'\;\Re\vph_\Sgm}}
 {1-\abs{C_0}^2\exp\brc{2(3k-2m)\cdot\int_0^y\dr y'\;
 \Re\vph_\Sgm}} = 0, 
 \label{eq_for_Sgm}
\ee
\be
 \vph^3 = 0, \;\;\;\;\;
 \vph^4 = \frac{C_0\exp\brc{\brkt{3k-2m}\cdot\int_0^y\dr y'\;
 \vph_\Sgm}}
 {\sbk{1-\abs{C_0}^2\exp\brc{2(3k-2m)\cdot\int_0^y\dr y'\;
 \Re\,\vph_\Sgm}}^{1/2}}, \;\;\;\;\;
 k\cdot\Im\,\vph_\Sgm = 0, 
 \label{eq_for_phis}
\ee
where $C_0$ is an integration constant. 
We have chosen the gauge where $\ge{y}{4}=1$.
Thus the $\bdm{D}$-gauge fixing condition~(\ref{D_gauge1}) becomes 
\be
 \cN(2\Re\,\vph_\Sgm)=1.  \label{bgd_norm}
\ee
Note that $\ge{y}{4}$ does not appear in the action~(\ref{action1}) 
explicitly, but appear only in the gauge fixing 
condition~(\ref{D_gauge1}) in our superfield description. 
The BPS background of $\vph_\Sgm^I$ is obtained 
as a solution of Eq.(\ref{eq_for_Sgm}). 
Plugging it into Eq.(\ref{eq_for_phis}), the background of $\vph^4$ 
is obtained. 

In the case that $\nV=0$ (and $C_{000}=1$), 
the BPS solution is 
\bea
 \sgm \eql -k_0 y+\frac{1}{3}\ln\frac{1-\abs{C_0}^2e^{(3k_0-2m_0)y}}
 {1-\abs{C_0}^2}, \nonumber\\
 \vph^3 \eql 0, \;\;\;\;\;
 \vph^4 = \frac{C_0\exp\brc{\brkt{\frac{3}{2}k_0-m_0}y}}
 {\sbk{1-\abs{C_0}^2\exp\brc{(3k_0-2m_0)y}}^{1/2}}, \;\;\;\;\;
 \vph_\Sgm^0 = \frac{1}{2}, 
\eea
for $0\leq y\leq \pi R$. 
Here we have chosen the gauge where $\ge{y}{4}=1$. 
Namely $R$ coincides with the radius of $S^1/Z_2$ in this case. 
In this model, $R$ is an arbitrary constant, which means that 
the radion mode is a modulus. 
The arbitrary constant~$C_0$ indicates the existence of another modulus  
in $\vph^4$. 
When $C_0$ is small, the background geometry is approximately 
the Randall-Sundrum warped spacetime, 
and the gauge couplings~$k_0$ and $m_0$ correspond to 
the $\mbox{AdS}_5$ curvature and the bulk (kink) mass for the hypermultiplet, 
respectively.

\subsection{$\nC=2$ case} \label{nc2_case}
Next we consider the two-compensator-hypermultiplet case. 
In this case, the manifold spanned by the hyperscalars becomes 
$SU(2,\nH)/SU(2)\times SU(\nH)\times U(1)$. 
Especially when $\nH=1$, this corresponds to the manifold of 
the universal hypermultiplet, which appears in the reduction 
of the heterotic M-theory on $S^1/Z_2$ to five dimensions. 
Because we have one more compensator multiplet in addition to the previous case, 
we introduce an abelian vector multiplet~$(V_T,\Sgm_T)$ 
without its own kinetic term to eliminate the additional 
degrees of freedom for the second compensator multiplet~\cite{KO2,FKO}. 
The charges of the hypermultiplets for this $U(1)_T$ vector multiplet 
are assigned as $igt_T=\sgm_3\otimes\bdm{1}_{2+\nH}$. 
Namely, $\Phi^{2\hat{a}-1}$ ($\Phi^{2\hat{a}}$) carries $+1$ ($-1$) charge. 
\begin{table}[t]
\begin{center}
\begin{tabular}{|c|c|c|c|c|c|c|c|c|c|c|} \hline
\rule[-2mm]{0mm}{7mm} & $V^{I'}$ & $\Sgm^{I'}$ & $V^{I''}$ & $\Sgm^{I''}$ 
 & $V_T$ & $\Sgm_T$ & $\Phi^1$ & 
 $\Phi^2$ & $\Phi^{2\hat{a}-1}$ & $\Phi^{2\hat{a}}$ 
 \\ \hline 
$Z_2$-parity & $-$ & $+$ & $+$ & $-$ & $+$ & $-$ & $-$ & $+$ & $+$ 
 & $-$ \\ \hline
\end{tabular}
\end{center}
\caption{The orbifold parities in the case of $\nC=2$. 
The indices run over $I'=0,1,\cdots,\nVp$; $I''=\nVp+1,\nVp+2,\cdots,\nV$;
$\hat{a}=2,3,\cdots,\nH+2$. }
\label{Z2_parity2}
\end{table}
According to Ref.~\cite{KO2,FKO}, we choose the $Z_2$-parities of 
the hypermultiplets as listed in Table~\ref{Z2_parity2}. 

Now we consider the boundary Lagrangians. 
Note that only $\Phi^2\Phi^3$ is gauge invariant for $U(1)_T$ 
among the combinations of the two $Z_2$-even chiral compensators. 
Thus $\cL^{(\vth)}_{\rm brane}$ has the following form. 
\bea
 \cL^{(\vth)}_{\rm brane} \eql \sbk{\int\dr^2\tht\;\frac{1}{4}
 f^{(\vth)}_{\bar{I}\bar{J}}(Q)\cW^{\bar{I}}\cW^{\bar{J}}+\hc}
 -3e^{2\sgm}\int\dr^4\tht\;\abs{\Phi^2\Phi^3}^{2/3}
 \exp\brc{-K^{(\vth)}(\bar{Q},Q,U)/3} \nonumber\\
 &&+e^{3\sgm}\sbk{\int\dr^2\tht\;(\Phi^2\Phi^3)P^{(\vth)}(Q)+\hc}. 
 \label{nc2_Lbd}
\eea
The powers of $\Phi^2\Phi^3$ are determined by the Weyl weight counting. 
Since the physical chiral multiplets must have zero Weyl weights and 
be neutral for $U(1)_T$, the hypermultiplets appear 
in $\cL^{(\vth)}_{\rm brane}$ in the form of 
\be
 Q^{\hat{a}} = \frac{\Phi^{2\hat{a}-1}}{\Phi^3}. \;\;\;\;\;
 (\hat{a}\geq 3)
\ee

As a specific example, we consider a case of $\nH=1$, 
where the system contains a single physical hypermultiplet, \ie, 
the universal hypermultiplet. 
The orbifold projection reduces the isometry group of 
the hyperscalar manifold~$U(2,1)$ to the subgroup~$U(1)\times U(1,1)$. 
We partially gauge this unbroken isometry 
by $Z_2$-odd vector multiplets $V^I$ ($I=I'$) as 
\be
 igt_I = \sgm_3\otimes \vep(y)
 \brkt{\begin{array}{ccc}-\bt_I&0&0\\0&\alp_I&\alp_I\\0&-\alp_I&-\alp_I
 \end{array}}, \label{nc2_igtI}
\ee
where $\alp_I$ and $\bt_I$ are gauge couplings 
for $U(1)$ and a subgroup of $U(1,1)$, respectively. 
This corresponds to a generalization of the situation in Ref.~\cite{LOSW} 
where an effective theory of the heterotic M theory is considered. 
The Lagrangian is written as 
\bea
 \cL \eql -3e^{2\sgm}\int\dr^4\tht\;\cN^{1/3}(\cV)\left\{e^{-2V_T}\brkt{
 e^{2\bt\cdot V}\abs{\Phi^1}^2+\bPhi^\dagger \sgm_3 
 e^{-2\alp\cdot V\tau}\bPhi}\right. \nonumber\\
 &&\hspace{40mm}\left.
 +e^{2V_T}\brkt{e^{-2\bt\cdot V}\abs{\Phi^2}^2+\bPhi^{c\dagger}\sgm_3
 e^{2\alp\cdot V\tau}\bPhi^c}\right\}^{2/3} \nonumber\\
 &&-2e^{3\sgm}\left[\int\dr^2\tht\;\left\{\Phi^1\brkt{
 \der_y+\frac{3}{2}\dot{\sgm}-2\bt\cdot\Sgm+2\Sgm_T}\Phi^2\right.\right.
 \nonumber\\
 &&\hspace{28mm}\left.\left.
 +\bPhi^t\sgm_3\brkt{\der_y+\frac{3}{2}\dot{\sgm}
 +2(\alp\cdot\Sgm)\tau+2\Sgm_T}\bPhi^c\right\}+\hc\right]+\cdots, 
 \label{tc_action1}
\eea
where the ellipsis denotes terms corresponding to the first line of 
(\ref{action2}), which are irrelevant to finding a BPS background, and 
$\alp\cdot V\equiv \vep(y)\alp_IV^I,\; 
\bt\cdot V\equiv \vep(y)\bt_IV^I,\; \cdots$,
\be
 \bPhi \equiv \vct{\Phi^3}{\Phi^5}, \;\;\;\;\;
 \bPhi^c \equiv \vct{\Phi^4}{\Phi^6}, \;\;\;\;\;
 \tau \equiv \sgm_3+i\sgm_2 = \mtrx{1}{1}{-1}{-1}. 
\ee
We have not introduced the boundary actions for simplicity. 

From the equations of motion for $V_T$ and $\Sgm_T$, we obtain 
\be
 e^{4V_T} = \frac{e^{2\bt\cdot V}\abs{\Phi^1}^2
 +\bPhi^\dagger \sgm_3 e^{-2(\alp\cdot V)\tau}\bPhi}
 {e^{-2\bt\cdot V}\abs{\Phi^2}^2+\bPhi^{c\dagger}\sgm_3
 e^{2(\alp\cdot V)\tau}\bPhi^c}, \label{EOM_VT}
\ee
\be
 \Phi^1\Phi^2+\bPhi^t\sgm_3\bPhi^c = 0. \label{EOM_SgmT}
\ee
Namely $\Sgm_T$ plays a role of a Lagrange multiplier. 
Using these equations, the action~(\ref{tc_action1}) becomes 
\bea
 \cL \eql -3e^{2\sgm}\int\dr^4\tht\;\brc{
 4\cN(\cV)\brkt{\abs{\Phi^1}^2+\bPhi^\dagger\sgm_3 
 e^{-2(\bt+\alp\tau)\cdot V}\bPhi}
 \brkt{\abs{\Phi^2}^2+\bPhi^{c\dagger}\sgm_3 e^{2(\bt+\alp\tau)\cdot V}
 \bPhi^c}}^{1/3} \nonumber\\
 &&-2e^{3\sgm}\sbk{\int\dr^2\tht\;\brc{\Phi^1\der_y\Phi^2
 +\bPhi^t\sgm_3\brkt{\der_y+2(\bt+\alp\tau)\cdot\Sgm}\bPhi^c}+\hc}
 +\cdots, 
 \label{tc_action2}
\eea
with the constraint~(\ref{EOM_SgmT}). 
Thus the BPS equations are read off as follows. 
\be
 \frac{e^{-2\sgm}}{\cN^{1/3}}\der_y\brc{e^{2\sgm}\frac{\cN_I}{\cN^{2/3}}}
 = 4\vep(y)\brc{\bvph^\dagger\sgm_3(\bt_I+\alp_I\tau)\bvph
 -\bvph^{c\dagger}\sgm_3(\bt_I+\alp_I\tau)\bvph^c}, \nonumber
\ee
\vspace{-10mm}
\bea
 \vep(y)\bvph^t\sgm_3(\bt_I+\alp_I\tau)\bvph^c \eql 0, \nonumber\\
 \brc{\der_y+3\dot{\sgm}-2(\bt+\alp\tau)\cdot\vph_\Sgm
 +\frac{\der_y\vph^2}{\vph^2}}\bvph \eql 0, \nonumber\\
 \brc{\der_y+2(\bt+\alp\tau)\cdot\vph_\Sgm
 -\frac{\der_y\vph^2}{\vph^2}}\bvph^c \eql 0, 
 \label{BPSeq2}
\eea
where $\bvph$ and $\bvph^c$ are the lowest components of 
$\bPhi$ and $\bPhi^c$, respectively. 
The arguments of $\cN$'s in the first equation are $2\Re\,\vph_\Sgm$. 
We have used the gauge-fixing condition~(\ref{T_gauge1}).  

Here we show a solution of (\ref{BPSeq2}) in the simplest case, \ie, 
$\nV=0$ and $C_{000}=1$. 
Let us choose the gauge where $\ge{y}{4}=1$. 
Then $\Re\,\vph_\Sgm^0=1/2$ from the $\bdm{D}$ 
gauge-fixing condition~(\ref{D_gauge1}). 
A BPS solution of (\ref{BPSeq2}) consistent 
with the gauge-fixing~(\ref{nc2_S_gauge}) is given for $0<y<\pi R$ as follows. 
\bea
 e^{6\sgm} \eql e^{2\bt_0 y}\brc{A+2B\alp_0y}, \;\;\;\;\;
 \Im\,\vph_\Sgm^0 = 0, \nonumber\\
 S \defa \frac{1-\phi_2}{1+\phi_2} = \frac{A}{B}+2\alp_0 y, \;\;\;\;\;
 \xi \equiv \frac{\phi_1}{1+\phi_2} = 0, \label{BPSsol1}
\eea
where $A$ and $B$ are real integration constants, and $\phi_1$, $\phi_2$ 
are the physical scalars defined in (\ref{nc2_S_gauge}). 
The real part of $S$ corresponds to the volume of the Calabi-Yau manifold 
in the heterotic M theory. 
When $\bt_0=0$, the solution~(\ref{BPSsol1}) agrees with that obtained 
in Ref.~\cite{LOSW} after moving to the gauge where 
$\ge{y}{4}=1$ in the latter. 
As can be seen from the result~(\ref{BPSsol1}), 
the Calabi-Yau volume~$\Re\,S$ depends on only $\alp_0$. 
The $\bt_0$-gauging affects on the background geometry and 
induces an exponential warp factor just like the Randall-Sundrum model. 
The linear dependence on $\alp_0$ in the above solution stems from 
the nilpotency of $\tau$. 

In the case of an arbitrary number of $\nV$, we can express 
a BPS solution in an analytic form if $\bt_I=0$ or $\alp_I=0$. 
As in the second paper of Ref.~\cite{LOSW}, 
the solution is expressed in terms of real functions~$f^I=f^I(y)$, 
which are implicitly defined as follows (for $0<y<\pi R$). 
When $\bt_I=0$, $f^I$ are defined by 
\be
 \cN_I(f) =8\alp_I By+c_I, \label{def_f1}
\ee
where $B$ and $c_I$ are real integration constants. 
We have chosen the gauge where $\ge{y}{4}=e^{4\sgm}$. 
So $R$ is no longer the radius of the orbifold. 
Using these functions, the BPS solution is expressed as 
\be
 e^{3\sgm} = \cN(f), \;\;\;\;\;
 \vph_\Sgm^I = \frac{e^{-\sgm}f^I}{2}, \;\;\;\;\;
 S = \frac{\cN^2(f)}{B}. 
\ee
When $\alp_I=0$, on the other hand, $f^I$ are defined by 
\be
 \cN_I(f) = 2\bt_I y+d_I, \label{def_f2}
\ee
where $d_I$ are real integration constants. 
We have chosen the gauge where $\ge{y}{4}=e^{-2\sgm}$. 
The BPS solution in this case is expressed as
\be
 e^{3\sgm} = \cN(f), \;\;\;\;\;
 \vph_\Sgm^I = \frac{e^{-\sgm}f^I}{2}, \;\;\;\;\;
 S = \mbox{(constant)}. 
\ee

\section{Off-shell dimensional reduction} \label{ODR}
In this section, we will derive the 4D effective action 
keeping the $N=1$ structure manifest. 
The key is a treatment of the $Z_2$-odd $N=1$ multiplets. 
Since they do not have zero modes, 
they should be integrated out from the low-energy effective action. 
As is shown in Ref.~\cite{LS,HLW}, heavy fields are eliminated by 
their equations of motion. 
Up to the two-derivative order for 4D spacetime coordinates~$x^\mu$, 
contributions from the kinetic terms for the heavy fields are 
negligible because they are suppressed by a power of 
$E_{\rm typ}/M_{\rm heavy}$, 
where $E_{\rm typ}$ is a typical energy scale in the effective theory 
and $M_{\rm heavy}$ is the mass scale of the heavy fields. 
Thus we can drop the kinetic terms for the heavy fields 
from the original action if we work up to the two-derivative order. 
The conformal SUGRA actions of Refs.~\cite{KO2,FKO,KO3}, 
on which our derivation is based, includes terms 
only up to two-derivative order. 
Therefore we will apply the following prescription 
to the 5D SUGRA action. 
\begin{description}
\item[Prescription:]
\be
 \mbox{\it Drop the kinetic terms for the $Z_2$-odd $N=1$ multiplets.}
 \label{prescription}
\ee
\end{description}
After this prescription, the $Z_2$-odd multiplets play similar roles 
to the Lagrange multipliers and provide constraints 
on the $Z_2$-even multiplets. 
In fact, such constraints extract the zero modes from the original 5D theory. 
As will be shown below, only 4D modes remain 
as the physical degrees of freedom after this prescription. 

\subsection{Extraction of 4D modes}
The action~(\ref{action2}) is invariant under the following 
gauge transformation up to total derivatives. 
\bea
 \tl{V}^I \eql V^I+\Lmd^I+\bar{\Lmd}^I, \nonumber\\
 \tl{\Sgm}^I \eql \Sgm^I+\der_y\Lmd^I, \nonumber\\
 \tl{\Phi}^a \eql \brkt{e^{2ig\Lmd^It_I}}^a_{\;\;b}\Phi^b, 
 \label{gauge_trf}
\eea
where the transformation parameters~$\Lmd^I$ ($I=0,1,\cdots,\nV$) 
are chiral multiplets. 
The (abelian) gauge groups act 
on each hypermultiplet~$(\Phi^{2\hat{a}-1},\Phi^{2\hat{a}})$ 
as $\sgm_3$ or $\sgm_{1,2}$. 
Since $\Phi^{2\hat{a}-1}$ and $\Phi^{2\hat{a}}$ have opposite $Z_2$-parities, 
$g\Lmd^I$ is even (odd) for the $\sgm_3$- ($\sgm_{1,2}$-) gauging. 

In this section we consider the $\sgm_3$-gauging case. 
The other case is discussed in Sect.~\ref{sgm1_gauging}. 
Let us choose the gauge transformation parameter~$\Lmd^I$ as 
\bea
 \Lmd^{I'}(y) \eql \Lmd_\Sgm^{I'}(y) \equiv 
 -\vep(y)\int_0^y\dr y'\;\vep(y')\Sgm^{I'}(y'), \nonumber\\
 \Lmd^{I''}(y) \eql \Lmd_\Sgm^{I''}(y) \equiv 
 -\int_0^y\dr y'\;\Sgm^{I''}(y').  
 \label{gLmd1}
\eea
Then $\Sgm^I$ are transformed to  
\bea
 \tl{\Sgm}^{I'} \eql \pi T^{I'}\dlt(y-\pi R)+\cdots, \nonumber\\
 \tl{\Sgm}^{I''} \eql 0,  \label{def_T}
\eea
where $T^{I'}$ are 4D chiral multiplets defined by 
\be
 T^{I'} \equiv \frac{2}{\pi}\int_0^{\pi R}\dr y\;\Sgm^{I'}(y). 
\ee
Thus $\Sgm^I$ are gauged away from the bulk Lagrangian. 
The ellipsis in (\ref{def_T}) denotes finite terms on the boundaries, 
which are neglected in the $y$-integral 
because their supports are measure zero. 
In the case of $\nVp=0$, $T^0$ corresponds to 
the so-called a radion multiplet. 
The first line of (\ref{action2}), $\cL_{\rm CS}$, corresponds to 
the supersymmetric Chern-Simons term and is invariant up to 
a total derivative for $y$ under the transformation~(\ref{gauge_trf}). 
Thus the following total derivative terms newly appear 
after rewriting the action in terms of the gauge-transformed quantities. 
\be
 \dlt\cL_{\rm CS} = \int\dr^2\tht\;
 \der_y\brkt{\frac{C_{IJK}}{2}\Lmd_\Sgm^I\tl{\cW}^J\tl{\cW}^K}+\hc. 
 \label{L_CS}
\ee
This becomes surface terms for the $y$-integration. 

In the following we consider a case that $(\nC,\nH)=(1,1)$ for simplicity 
while $\nV$ is left to be an arbitrary number. 
Since $\tl{\Phi}^1$ is $Z_2$-odd, the 4D chiral compensator multiplet 
is expected to be contained in $\tl{\Phi}^2$. 
Thus we redefine the hypermultiplet~$\tl{\Phi}^a$ ($a=1,2,3,4$) as 
\bea
 \tl{\Phi}^1 \eql e^{-\frac{3}{2}\sgm}\phi^{\frac{3}{2}}\phi^c, 
 \;\;\;\;\;
 \tl{\Phi}^2 = e^{-\frac{3}{2}\sgm}\phi^{\frac{3}{2}}, \nonumber\\
 \tl{\Phi}^3 \eql e^{-\frac{3}{2}\sgm}\phi^{\frac{3}{2}}H^c, \;\;\;\;\;
 \tl{\Phi}^4 = e^{-\frac{3}{2}\sgm}\phi^{\frac{3}{2}}H. 
 \label{fctout_phi}
\eea
We have rescaled $\tl{\Phi}^a$ by the warp factor 
so that explicit $\sgm$-dependence disappears from the action. 
Only $\phi$ carries a nonzero Weyl weight, \ie, one, 
and the other multiplets~$(\phi^c,H^c,H)$ have zero weights. 
Then the Lagrangians~(\ref{action2}) and (\ref{L_bd1}) are rewritten 
as the following form. 
\bea
 \cL \eql \sbk{\int\dr^2\tht\;\brc{\frac{C_{IJK}}{8}\bar{D}^2
 \brkt{\tl{V}^ID^\alp\der_y\tl{V}^J-D^\alp\tl{V}^I\der_y\tl{V}^J}
 \tl{\cW}^K_\alp}+\hc}+\dlt\cL_{\rm CS} 
 \nonumber\\
 &&-3\int\dr^4\tht\;\cN^{1/3}(\cV)\abs{\phi}^2
 \brc{e^{-3k\cdot\tl{V}}\abs{\phi^c}^2+e^{3k\cdot\tl{V}}
 -e^{-2m\cdot\tl{V}}\abs{H^c}^2-e^{2m\cdot\tl{V}}\abs{H}^2}^{2/3} \nonumber\\
 &&-2\sbk{\int\dr^2\tht\;\brc{\phi^{\frac{3}{2}}\phi^c
 \der_y(\phi^{\frac{3}{2}})-\phi^{\frac{3}{2}}H^c\der_y
 (\phi^{\frac{3}{2}}H)}+\hc} \nonumber\\
 &&+\sum_{\vth=0,\pi}\sbk{\int\dr^2\tht\;
 \phi^3P^{(\vth)}(Q)+\hc}\dlt(y-\vth R),  
 \nonumber\\
 \label{action3}
\eea
where the hypermultiplets are charged for $V^{I}$ as 
\be
 igt_{I'} = \sgm_3\otimes\vep(y)\dgnl{\frac{3}{2}k_{I'}}{m_{I'}}, \;\;\;\;\;
 igt_{I''} = \sgm_3\otimes\dgnl{\gc_{I''}}{\gh_{I'}}, 
 \label{sgm3_igtI}
\ee
and $\frac{3}{2}k\cdot\tl{V}\equiv\frac{3}{2}\vep(y)k_{I'}\tl{V}^{I'}
+\gc_{I''}\tl{V}^{I''}$, 
$m\cdot\tl{V}\equiv\vep(y)m_{I'}\tl{V}^{I'}+\gh_{I''}\tl{V}^{I''}$. 
For simplicity, we have introduced 
only the superpotentials~$P^{(\vth)}$ in the boundary action. 

Now we perform the prescription~(\ref{prescription}). 
Specifically, it is translated into the following. 
\begin{itemize}
 \item Drop the $Z_2$-odd chiral multiplets in the $d^4\tht$-integral.
 \item Drop the $Z_2$-odd superfield strengths~$\cW_\alp^{I'}$. 
\end{itemize}
Then $\phi^c$ and $H^c$ appear only in the $d^2\tht$-integral 
in the third line of (\ref{action3}). 
From their equations of motion, we obtain 
\be
 \der_y\phi=0, \;\;\;\;\;
 \der_yH=0. 
 \label{4Dchiral_cond}
\ee
This means that $\phi$ and $H$ are 4D multiplets. 
Hence $\phi$ is identified with the 4D chiral compensator. 
Note that the constraints in (\ref{4Dchiral_cond}) are obtained 
independently of the background, in contrast to Ref.~\cite{PST2} 
where they are obtained from the BPS equations. 

$\Sgm^I$ correspond to the gauge fields that covariantize the derivative 
operator~$\der_y$. 
Thus Eq.(\ref{4Dchiral_cond}) restricts the allowed gauge transformation 
to the one which keeps $\tl{\Sgm}^I=0$ 
because the derivative operator~$\der_y$  
acting on the hypermultiplets are now absent in the action. 
Namely only the $y$-independent $\Lmd^{I''}$ are allowed.\footnote{
Nonvanishing $y$-independent $\Lmd^{I'}$ are not allowed 
since they are $Z_2$-odd. } 
This indicates that the corresponding gauge multiplets~$V^{I''}$ 
must also be independent of $y$. 
\be
 \der_y\tl{V}^{I''}=0. \label{4Dvector_cond}
\ee 
Using this, the first line of (\ref{action3}), $\cL_1$, 
becomes a total derivative for $y$. 
\bea
 \cL_1 \defa \int\dr^2\tht\;\brc{\frac{C_{I''J'K''}}{8}\bar{D}^2\brkt{
 \tl{V}^{I''}D^\alp\der_y\tl{V}^{J'}-D^\alp\tl{V}^{I''}\der_y\tl{V}^{J'}}
 \tl{\cW}^{K''}_\alp}+\hc+\dlt\cL_{\rm CS} \nonumber\\
 \eql \int\dr^2\tht\;\der_y\left\{\frac{C_{I''J'K''}}{8}\bar{D}^2\brkt{
 \tl{V}^{I''}D^\alp\tl{V}^{J'}-D^\alp\tl{V}^{I''}\tl{V}^{J'}}
 \tl{\cW}^{K''}_\alp \right. \nonumber\\
 &&\hspace{20mm}\left.
 +\frac{C_{I'J''K''}}{2}\Lmd_\Sgm^{I'}\tl{\cW}^{J''}\tl{\cW}^{K''}
 \right\}+\hc.  \label{def_L1}
\eea
This becomes surface terms for the $y$-integration. 
Here note that the transformed quantities by $\Lmd_\Sgm$ are 
discontinuous or singular at $y=\pi R$. 
(See Eq.(\ref{def_T}).)
This singularity is an artifact due to the discontinuous 
gauge transformation. 
The bulk Lagrangian has no singular terms in the original field basis. 
So we can safely neglect the contributions on the boundary~$y=\pi R$ 
for the bulk action.  
In such a case, the above surface terms have nonvanishing values at 
$y\to\pi R-$ while there are no contributions from the delta functions 
in (\ref{def_T}). 
On the other hand, if we take the domain of the $y$-integration as 
$0\leq y\leq \pi R$, the surface terms vanishes at $y=\pi R$ 
while the delta functions in (\ref{def_T}) 
compensate their contributions, 
and the same result is obtained as the case of $0\leq y<\pi R$. 
Thus, in the following, 
we will take the domain of the $y$-integration as $0\leq y<\pi R$ 
for the bulk action 
in order to avoid the singular contributions at $y=\pi R$. 
From Eqs.(\ref{gLmd1}) and (\ref{def_T}), 
\be
%
 \lim_{y\to\pi R-}\Lmd_\Sgm^{I'} = -\frac{\pi}{2}T^{I'},  \label{prpty_LmdSgm}
\ee
and thus 
\be
%
 \lim_{y\to \pi R-}\tl{V}^{I'} = -\pi\Re\,T^{I'}. \label{lim_VIp}
\ee
We have assumed the continuity of $V^{I'}$ at the boundaries. 
Using (\ref{prpty_LmdSgm}) and (\ref{lim_VIp}), the 4D effective Lagrangian 
coming from $\cL_1$ is calculated as 
\be
 \cL^{\rm (4D)}_1 \equiv \int_0^{\pi R-}\!\!\!\!dy\;\cL_1 =
%
%
 -\int\dr^2\tht\;\frac{3\pi}{4}C_{I''J'K''}T^{J'}\tl{\cW}^{I''}\tl{\cW}^{K''}
 +\hc+\cdots. 
\ee
We have used the relations~$d^2\btht=-\frac{1}{4}\bar{D}^2$, 
$D^\alp\tl{\cW}_\alp=\bar{D}_{\dalp}\bar{\tl{\cW}}^{\dalp}$ and 
the partial integral for $\tht^\alp$ in the calculation. 
The ellipsis denotes terms involving $D_\alp T$ or 
$\bar{D}_{\dalp}\bar{T}$. 
They involve higher order derivative terms for the 4D coordinates~$x^\mu$ 
and are dropped in a truncation at two-derivative order we are working. 

There is an important point to notice about Eq.(\ref{4Dchiral_cond}). 
It holds only in the bulk ($0<y<\pi R$) because it comes from 
the equations of motion for the $Z_2$-odd multiplets, which vanish 
on the boundaries. 
Thus $\phi$ and $H$ can have different values on the boundaries 
from their bulk values~$\phi_{\rm bulk}$ and $H_{\rm bulk}$. 
In fact, from Eqs.(\ref{gauge_trf}), (\ref{prpty_LmdSgm}) and 
$\vep(\pi R)=0$, 
\bea
 \lim_{y\to\pi R-}\tl{\Phi}^2(y) \eql
 e^{\frac{3}{2}\pi k_{I'}T^{I'}}e^{-2\gc_{I''}\Lmd_\Sgm^{I''}}\Phi^2(\pi R) 
 = e^{\frac{3}{2}\pi k_{I'}T^{I'}}\tl{\Phi}^2(\pi R), \nonumber\\
 \lim_{y\to\pi R-}\tl{\Phi}^4(y) \eql
 e^{\pi m_{I'}T^{I'}}e^{-2\gh_{I''}\Lmd_\Sgm^{I''}}\Phi^4(\pi R) 
 = e^{\pi m_{I'}T^{I'}}\tl{\Phi}^4(\pi R).
\eea
From this and (\ref{fctout_phi}), we can see 
\bea
 \phi_{\rm bulk} \eql e^{\pi k_{I'}T^{I'}}\phi|_{\pi R}, \nonumber\\
 H_{\rm bulk} \eql e^{-\pi\brkt{\frac{3}{2}k_{I'}-m_{I'}}T^{I'}}H|_{\pi R}. 
 \label{rel_phis1}
\eea
On the other hand, $\phi$ and $H$ are continuous at $y=0$ because 
$\Lmd_\Sgm^I$ are continuous there. 
\be
 \phi_{\rm bulk} = \phi|_0, \;\;\;\;\;
 H_{\rm bulk} = H|_0. 
\ee

The 4D effective Lagrangian~$\cL^{\rm (4D)}$ is obtained 
by performing the $y$-integration for the 5D Lagrangian. 
\bea
 \cL^{\rm (4D)} \eql -\sbk{\int\dr^2\tht\;\frac{3\pi}{4}C_{I'J''K''}T^{I'}
 \tl{\cW}^{J''}\tl{\cW}^{K''}+\hc} \nonumber\\
 &&-3\int\dr^4\tht\;\abs{\phi}^2\brc{\int_0^{\pi R-}\!\!\!\!dy\;
 \hat{\cN}^{1/3}(-\der_y\tl{V})\,
 e^{2k\cdot\tl{V}}\brkt{1-e^{-(3k-2m)\cdot\tl{V}}\abs{H}^2}^{2/3}}
 \nonumber\\
 &&+\sbk{\int\dr^2\tht\;\phi^3\frac{1}{2}\brc{P^{(0)}(Q)
 +e^{-3\pi k_{I'}T^{I'}}P^{(\pi)}(Q)}+\hc}, \nonumber\\
 \label{red_action1}
\eea
where $\phi$ and $H$ are understood as $\phi_{\rm bulk}$ and $H_{\rm bulk}$, 
and 
\be
 \hat{\cN}(X) \equiv C_{I'J'K'}X^{I'}X^{J'}X^{K'}. 
\ee
From Eqs.(\ref{nc1_US}), (\ref{fctout_phi}), (\ref{rel_phis1}) 
and $\vep(\vth R)=0$ ($\vth=0,\pi$),  
\be
 Q = \left.\frac{\Phi^4}{\Phi^2}\right|_{\vth R} 
 =H|_{\vth R} =\exp\brc{\vth\brkt{\frac{3}{2}k_{I'}-m_{I'}}T^{I'}}
 H_{\rm bulk}
\ee
in the case that $Q$ is an induced multiplet from the bulk. 
Here we have dropped $\Lmd_\Sgm^{I''}$-dependence in the boundary actions 
because they will be cancelled out due to the gauge invariance. 

The action~(\ref{red_action1}) contains all the 4D multiplets 
that should appear in the effective theory. 
The only task to derive the effective Lagrangian is integrating out 
$\tl{V}^{I'}$ in (\ref{red_action1}).

\subsection{Integrating out $\tl{V}^{I'}$} \label{VI_integrate}

\subsubsection{$\nVp=0$ case} \label{nVp0_case}
As pointed out in Ref.~\cite{PST2}, $\tl{V}^{I'=0}$ can easily be integrated out 
in the case of $\nVp=0$. 
Since $\hat{\cN}^{1/3}(-\der_y\tl{V})=-\der_y\tl{V}^0$ in this case, 
the $y$-integral in (\ref{red_action1}) can be performed straightforwardly. 
\bea
 &&\int_0^{\pi R-}\!\!\!\! dy\;\hat{\cN}^{1/3}(-\der_y\tl{V})e^{2k\cdot\tl{V}}
 \brkt{1-e^{-(3k-2m)\cdot\tl{V}}\abs{H}^2}^{2/3} \nonumber\\
 \eql -\int_0^{-\pi\Re\,T}\!\!d\tl{V}^0\;e^{2k\tl{V}^0}
 \brkt{1-e^{-(3k-2m)\tl{V}^0}\abs{H}^2}^{2/3} \nonumber\\
 \eql \frac{1-e^{-2\pi k\Re\,T}}{2k}-\frac{2\brkt{e^{\pi(k-2m)\Re\,T}-1}}
 {3(k-2m)}\abs{H}^2+\cO\brkt{\abs{H}^4},  
\eea
where $k\equiv k^0$, $m\equiv m^0$, and 
$T\equiv T^0$ corresponds to the radion multiplet. 
We have used (\ref{lim_VIp}). 
Note that we do not need to know an explicit $y$-dependence of $\tl{V}^0$ 
in order to derive the 4D effective Lagrangian. 
In fact, the $y$-dependence of $\tl{V}^0$ remains undetermined 
in the off-shell dimensional reduction. 
The situation is similar in the case of $\nVp\geq 1$. 
We can derive the 4D effective Lagrangian without the knowledge about 
explicit $y$-dependences of $\tl{V}^{I'}$ as will be shown in the folowing. 

\subsubsection{$\nVp\geq 1$ case}
The equations of motion for $\tl{V}^{I'}$ are obtained 
from (\ref{red_action1}) as  
\bea
 &&\hat{\cN}^{1/3}e^{2k\cdot\tl{V}}\brc{
 2k_{I'}\brkt{1-e^{-(3k-2m)\cdot\tl{V}}\abs{H}^2}^{2/3} 
 +\frac{2}{3}\frac{
 (3k_{I'}-2m_{I'})e^{-(3k-2m)\cdot\tl{V}}\abs{H}^2}
 {\brkt{1-e^{-(3k-2m)\cdot\tl{V}}\abs{H}^2}^{1/3}}} \nonumber\\
 &&-\der_y\brc{-\frac{\hat{\cN}_{I'}}{3\hat{\cN}^{2/3}}
 e^{2k\cdot\tl{V}}\brkt{1-e^{-(3k-2m)\cdot\tl{V}}\abs{H}^2}^{2/3}} =0, 
 \label{EOM_Vp1}
\eea
for $0<y<\pi R$. 
Here and henceforth, the arguments of $\hat{\cN}$'s are understood 
as $\cV^{I'}=-\der_y\tl{V}^{I'}$ unless they are explicitly specified. 
This can be rewritten as
\be
 \brc{\der_y\brkt{\frac{\hat{\cN}_{J'}}{\hat{\cN}}}
 +6k_{J'}+\frac{2(3k_{J'}-2m_{J'})e^{-(3k-2m)\cdot\tl{V}}\abs{H}^2}
 {1-e^{-(3k-2m)\cdot\tl{V}}\abs{H}^2}}\cP^{J'}_{\;\;I'}(-\der_y\tl{V}) 
 = 0. \label{EOM_Vp2}
\ee
where 
\be
 \cP^{J'}_{\;\;I'}(X) \equiv \dlt^{J'}_{\;\;I'}-\frac{X^{J'}\hat{\cN}_{I'}}
 {3\hat{\cN}}(X)  
\ee
is a projection operator which satisfies 
\be
 \cP^{J'}_{\;\;I'}(X)X^{I'} = \hat{\cN}_{J'}(X)\cP^{J'}_{\;\;I'}(X) = 0. 
\ee
Thus only $\nVp$ equations among $(\nVp+1)$ ones 
are independent in Eq.(\ref{EOM_Vp2}). 

To illustrate the procedure, 
we consider a special case that $3k_{I'}=2m_{I'}$ for all $I'$ 
in this subsection.\footnote{ 
The basic procedure is similar also 
in a generic case that $3k_{I'}\neq 2m_{I'}$ for some $I'$, 
which is discussed in Appendix~\ref{3kI_neq_2mI}. } 
Namely the compensator and the physical hypermultiplets have 
the same charges for all $V^{I'}$.
In this case, Eq.(\ref{EOM_Vp1}) is reduced to 
\be
 \hat{\cN}^{1/3}e^{2k\cdot\tl{V}}
 \brkt{1-e^{2(\gh-\gc)_{I''}\tl{V}^{I''}}\abs{H}^2}^{2/3}
 = -\frac{1}{2k_{\check{I}'}}
 \der_y\brc{\frac{\hat{\cN}_{\check{I}'}}{3\hat{\cN}^{2/3}}e^{2k\cdot\tl{V}}
 \brkt{1-e^{2(\gh-\gc)_{I''}\tl{V}^{I''}}\abs{H}^2}^{2/3}}. 
 \label{EOM_Vp3}
\ee
The checked index~$\check{I}'$ denotes that we do not sum over it. 
Note that the left-hand side is equal to the integrand of 
the $d^4\tht$-integral in (\ref{red_action1}).  
Thus Eq.(\ref{EOM_Vp3}) means that the second line of (\ref{red_action1}) 
can be rewritten as surface terms for the $y$-integral. 

Since $\hat{\cN}$ is a cubic polynomial, 
$\hat{\cN}_{I'}/\hat{\cN}^{2/3}$ are functions of 
$\nVp$ independent variables~$v^i$ ($i=1,2,\cdots,\nVp$) defined by 
\be
 v^i \equiv \frac{\der_y\tl{V}^{I'=i}}{\der_y\tl{V}^0}. 
 \label{def_vi}
\ee
Namely, 
\be
 \frac{\hat{\cN}_{\check{I}'}}{2k_{\check{I}'}
 \hat{\cN}^{2/3}}(-\der_y\tl{V}) \equiv \cF_{I'}(v). 
 \label{def_cF}
\ee
Using (\ref{EOM_Vp3}) and (\ref{def_cF}), the 4D Lagrangian~(\ref{red_action1})  
can be rewritten as 
\bea
 \cL^{\rm (4D)} \eql -\sbk{\int\dr^2\tht\;\frac{3\pi}{4}C_{I'J''K''}T^{I'}
 \tl{\cW}^{J''}\tl{\cW}^{K''}+\hc} \nonumber\\
 &&+\int\dr^4\tht\;\abs{\phi}^2
 \brc{\cF_{J'}(v_\pi)e^{-2\pi k_{I'}\Re\,T^{I'}}
 -\cF_{J'}(v_0)}\brkt{
 1-e^{2(\gh-\gc)_{I''}\tl{V}^{I''}}\abs{H}^2}^{2/3} \nonumber\\
 &&+\sbk{\int\dr^2\tht\;\phi^3\frac{1}{2}\brc{P^{(0)}(Q)
 +e^{-3\pi k_{I'}T^{I'}}P^{(\pi)}(Q)}+\hc}, 
 \label{4D_L1}
\eea
for each $J'$. 
Here $v^i_0$ and $v^i_\pi$ are 
boundary values of $v^i$ at $y=0$ and $y=\pi R$, respectively. 

The last task to obtain the effective Lagrangian is to express 
$v^i_0$ and $v^i_\pi$ in terms of the physical multiplets. 
Since the $H$-dependence of Eq.(\ref{EOM_Vp3}) can be factored out~\footnote{
Recall that both $H$ and $V^{I''}$ are $y$-independent.}, 
Eq.(\ref{EOM_Vp3}) is rewritten as 
\be
 \der_y\brkt{\frac{\hat{\cN}_{\check{I}'}}{k_{\check{I}'}\hat{\cN}^{2/3}}} 
 = -\frac{6(k\cP)_{\check{I}'}}{k_{\check{I}'}}\hat{\cN}^{1/3}, 
 \label{EOM_Vp4}
\ee
where $(k\cP)_{I'}\equiv k_{J'}\cP^{J'}_{\;\;I'}$, 
and the argument of $\cP$ is $-\der_y\tl{V}$.  
The right-hand side of (\ref{EOM_Vp4}) can be rewritten 
in the following form. 
\be
 \mbox{(R.H.S.)} = -\cG_{I'\check{J}'}(v)\der_y\tl{V}^{\check{J}'}, 
\ee
where $\cG_{I'J'}$ are some functions of $v^i$. 
Thus, from Eq.(\ref{EOM_Vp4}), we obtain 
\be
 -\der_y\tl{V}^{J'} = \frac{\der_y\cF_{\check{I}'}(v)}
 {\cG_{\check{I}'J'}(v)} 
 \equiv \sum_{i=1}^{\nVp}\cH_{I'J'i}(v^1,v^2,\cdots,v^{\nVp})\der_yv^i, 
 \label{der_yV1}
\ee
where $J'=0,1,\cdots,\nVp$. 
Since the left-hand side is independent of $I'$, there are 
$(\nVp-1)$ independent equations~\footnote{
Recall that only $\nVp$ equations among $(\nV+1)$ ones 
in (\ref{EOM_Vp4}) are independent. 
Thus $I'$ runs over only $\nVp$ different values in (\ref{der_yV1}). }
among $\cH_{I'J'i}(v)$ for given 
$J'$ and $i$. 
\be
 \cH_{1J'i}(v) = \cH_{2J'i}(v) = \cdots = \cH_{\nVp J'i}(v). 
\ee
Using these equations, we can rewrite the coefficient function~$\cH_{1J'i}(v)$ 
so that it depends on only $v^i$ for each $i$. 
Namely, (\ref{der_yV1}) for $I'=1$ is rewritten as the following form. 
\be
 -\der_y\tl{V}^{J'} = \sum_{i=1}^{\nVp}\hat{\cH}_{J'i}(v^i)\der_yv^i. 
\ee
Integrating this over $[0,\pi R)$, we obtain 
\be
 \pi\Re\,T^{J'} = \sum_i\int_{v_0^i}^{v_\pi^i}\dr v^i\;
 \hat{\cH}_{J'i}(v^i) = \sum_i\brc{\cQ_{J'i}(v_\pi^i)-\cQ_{J'i}(v_0^i)},  
 \label{piT}
\ee
where $\cQ_{J'i}(v^i)$ are integrals of $\hat{\cH}_{J'i}(v^i)$. 
From the second line of (\ref{4D_L1}), we obtain 
the following $(\nVp-1)$ independent equations 
among $(v_0^i,v_\pi^i)$ and $T^{I'}$.\footnote{
Since only $\nVp$ equations are independent in Eq.(\ref{EOM_Vp3}), 
the number of independent equations is $(\nVp-1)$. }
\bea
 \cF_1(v_\pi)e^{-2k_{I'}\Re\,T^{I'}}-\cF_1(v_0) 
 \eql \cF_2(v_\pi)e^{-2k_{I'}\Re\,T^{I'}}-\cF_2(v_0)  \nonumber\\
 \eql \cdots 
 = \cF_{\nVp}(v_\pi)e^{-2k_{I'}\Re\,T^{I'}}-\cF_{\nVp}(v_0). 
 \label{FFs1}
\eea
Using these equations, we can express $(\nVp-1)$ variables, 
say $(v_\pi^2,\cdots,v_\pi^{\nVp})$, 
in terms of the other $v$'s and $\Re\,T^{I'}$. 
\bea
 v_\pi^2 \eql v_\pi^2(v_0^1,v_0^2,\cdots,v_0^{\nVp},v_\pi^1,\Re\,T^0,
 \Re\,T^1,\cdots,\Re\,T^{\nVp}), \nonumber\\
  &\vdots& \nonumber\\
 v_\pi^{\nVp} \eql v_\pi^{\nVp}(v_0^1,v_0^2,\cdots,v_0^{\nVp},v_\pi^1,\Re\,T^0,
 \Re\,T^1,\cdots,\Re\,T^{\nVp}).  \label{v2_to_vnVp}
\eea
Then Eq.(\ref{piT}) becomes the following form. 
\be
 \pi\Re\,T^{J'} = \cT^{J'}(v_0^1,v_0^2,\cdots,v_0^{\nVp},v_\pi^1,\Re\,T^0,
 \Re\,T^1,\cdots,\Re\,T^{\nVp}), 
 \label{piT2}
\ee
where $J'=0,1,\cdots,\nVp$. 
Solving these equations for $(v_0^1,v_0^2,,\cdots,v_0^{\nVp},v_\pi^1)$ 
and substituting them into Eq.(\ref{v2_to_vnVp}), 
all $(v_0^i,v_\pi^i)$ are expressed in terms of $\Re\,T^{I'}$. 
\bea
 v_0^i \eql v_0^i(\Re\,T^0,\Re\,T^1,\cdots,\Re\,T^{\nVp}), \nonumber\\
 v_\pi^i \eql v_\pi^i(\Re\,T^0,\Re\,T^1,\cdots,\Re\,T^{\nVp}), 
 \label{vis}
\eea
where $i=1,2,\cdots,\nVp$. 
Plugging these into (\ref{4D_L1}), we obtain the final result. 

In the case that $3k_{I'}\neq 2m_{I'}$ for some $I'$, 
we can perform a similar procedure and obtain the 4D effective action 
as shown in Appendix~\ref{3kI_neq_2mI}.

\subsection{More practical method} \label{practical}
Note that the derivation of the 4D effective action in the previous subsections 
does not need any information about the background. 
Namely we can perform the procedure regardless of whether 
the background is supersymmetric or not. 
This is because we keep the $N=1$ off-shell structure 
during the process. 
The background information is obtained only after we move to 
the on-shell action. 
However, for $\nVp\geq 1$, solving (\ref{piT2}) to obtain (\ref{vis}) 
is generically a hard task if $\hat{\cN}$ is a polynomial. 
Thus we propose alternative method to obtain the effective action. 

Although the fact that the 4D action can be derived without determining 
the explicit $y$-dependences of $\tl{V}^{I'}$ is important, 
the derivation of the 4D action becomes easier if we know them. 
Eq.(\ref{EOM_Vp2}) means that 
\be
 \der_y\brkt{\frac{\hat{\cN}_{I'}}{\hat{\cN}}}+6k_{I'}
 +\frac{2(3k_{I'}-2m_{I'})e^{-(3k-2m)\cdot\tl{V}}\abs{H}^2}
 {1-e^{-(3k-2m)\cdot\tl{V}}\abs{H}^2}
 = A\hat{\cN}_{I'}, 
 \label{EOM_Vp5}
\ee
where $A=A(\tl{V},\der_y\tl{V},H)$ is an arbitrary function. 
The funciton~$A$ is undetermined in the off-shell dimensional reduction. 
Once its explicit form is given, however, explicit $y$-dependences of 
$\tl{V}^{I'}$ are determined. 
As is clear from the discussion in Sect.~\ref{VI_integrate}, 
the final result does not depend on the explicit function form of $A$. 
Therefore we can assume some definite function form of $A$ by hand 
in order to calculate the 4D action. 
It is convenient to choose A as zero. 
Namely, 
\be
 \der_y\brkt{\frac{\hat{\cN}_{I'}}{\hat{\cN}}}+6k_{I'}
 +\frac{2(3k_{I'}-2m_{I'})e^{-(3k-2m)\cdot\tl{V}}\abs{H}^2}
 {1-e^{-(3k-2m)\cdot\tl{V}}\abs{H}^2} = 0. 
 \label{EOM_with_A0}
\ee
Multiplying this by $\der_y\tl{V}^{I'}$ and contracting $I'$, 
we obtain 
\be
 \der_y\brc{\hat{\cN}e^{6k\cdot\tl{V}}\brkt{1-e^{-(3k-2m)\cdot\tl{V}}
 \abs{H}^2}^2}=0. 
\ee
This means that the integrand in the second line of (\ref{red_action1}) 
is independent of $y$. 
Therefore the 4D effective Lagrangian is obtained by taking the integrand 
as that of $y=0$. 
\bea
 \cL^{\rm (4D)} \eql -\sbk{\int\dr^2\tht\;\frac{3\pi}{4}C_{I'J''K''}T^{I'}
 \tl{\cW}^{J''}\tl{\cW}^{K''}+\hc} \nonumber\\
 &&-3\pi R\int\dr^4\tht\;\abs{\phi}^2\left.\hat{\cN}^{1/3}\right|_{y=0}
 \brkt{1-e^{2(\gh-\gc)_{I''}\tl{V}^{I''}}\abs{H}^2}^{2/3} \nonumber\\
 &&+\sbk{\int\dr^2\tht\;\phi^3\frac{1}{2}
 \brc{P^{(0)}(Q)+e^{-3\pi k_{I'}T^{I'}}P^{(\pi)}(Q)}+\hc}, 
 \label{4D_L_practical}
\eea
where $\hat{\cN}^{1/3}|_{y=0}$ is evaluated by using a solution of 
(\ref{EOM_with_A0}). 
The equivalence of (\ref{4D_L_practical}) and (\ref{4D_L1}) is trivial 
because both are derived from the same expression~(\ref{red_action1}). 

When the gauge~$e_y^{\;\;4}=e^{-2\sgm}$ is chosen, 
Eq.(\ref{eq_for_Sgm}) becomes 
\be
 \der_y\brkt{e^{6\sgm}\cN_I(2\Re\,\vph_\Sgm)}+6k_I
 +\frac{2(3k_I-2m_I)\abs{C_0}^2
 \exp\brc{2(3k-2m)\cdot\int_0^y\dr y'\;\Re\vph_\Sgm}}
 {1-\abs{C_0}^2\exp\brc{2(3k-2m)\cdot\int_0^y\dr y'\;
 \Re\vph_\Sgm}} = 0. 
\ee
This has the same form as Eq.(\ref{EOM_with_A0}) 
under the following replacement.\footnote{
In Sect.~\ref{nc1_case}, $\cN=\hat{\cN}$ and $I=I'$ 
since all~$V^I$ are assumed to be $Z_2$-odd. } 
\bea
 \hat{\cN}(-\der_y\tl{V}) & \to & e^{-6\sgm}, \nonumber\\
 -\der_y\tl{V}^{I'} & \to & 2\Re\,\vph_\Sgm^{I'}, \nonumber\\
 H & \to & C_0. 
\eea
In the case that $\nVp\geq 1$, the BPS solution~$\vph_\Sgm^{I'}$ generically 
has integration constants, just like $c_I$ in (\ref{def_f1}) 
in the case of $\nC=2$. 
Such integration constants (in terms of $y$) in the solution of 
(\ref{EOM_with_A0}) corresponds to $\Re\,T^{I'}$. 
Here we have used the fact that the 4D K\"{a}hler potential is 
independent of $\Im\,T^{I'}$, which has been shown in the previous subsection. 
Therefore we can use a BPS solution to derive the 4D effective action 
in the case discussed in Sect.~\ref{nc1_case}. 
This corresponds to a method proposed in Ref.~\cite{PST2}. 
In fact, we can check that the off-shell 4D actions in Ref.~\cite{PST2} 
are reproduced by the method in the previous subsection 
in the case that $\hat{\cN}$ is a monomial. 
However, there is one important point to notice. 
Our method of the off-shell dimensional reduction can always be performed 
regardless of whether the background is supersymmetric or not. 
Eq.(\ref{EOM_with_A0}) always have a solution even if the background is 
non-supersymmetric while the BPS equations do not. 
Thus the correspondence between (\ref{EOM_with_A0}) 
and the BPS equations is broken in such a case. 
In fact, even in a supersymmetric case, 
the correspondence can be broken in some models. 
We will show such an example in Sect.~\ref{MOmodel}.

\section{Specific examples} \label{sp_ex}
In this section, we consider some specific models and demonstrate 
the procedure of deriving the 4D effective action 
discussed in the previous section. 

\subsection{Tadpole boundary superpotentials} \label{MOmodel}
Here we introduce the following boundary superpotentials 
to the model discussed in Sect.~\ref{nc1_case}. 
\be
 P^{(0)} = J_0\frac{\Phi^4}{\Phi^2}, \;\;\;\;\;
 P^{(\pi)} = -J_\pi\frac{\Phi^4}{\Phi^2}, \label{P_vth}
\ee
where $J_0$ and $J_\pi$ are positive constants. 

The BPS equations are read off from the action as in Sect.~\ref{nc1_case}. 
Due to the existence of the boundary terms, 
the fourth and sixth equations in (\ref{BPSeq1}) are modified as 
\bea
 \brkt{\der_y+\frac{3}{2}\dot{\sgm}-3k\cdot\vph_\Sgm}\vph^1 \eql 
 -\frac{\vph^4}{2}\brc{J_0\dlt(y)-J_\pi\dlt(y-\pi R)}, \nonumber\\
 \brkt{\der_y+\frac{3}{2}\dot{\sgm}-2m\cdot\vph_\Sgm}\vph^3 \eql 
 \frac{\vph^2}{2}\brc{J_0\dlt(y)-J_\pi\dlt(y-\pi R)}. 
 \label{BPSeq_MOmodel}
\eea
Using the gauge-fixing condition~(\ref{nc1_S_gauge}), we can see 
from the first equation in (\ref{BPSeq_MOmodel}) 
that $\vph^4$ cannot have nonvanishing background value. 
On the other hand, $\vph^3$ has a nontrival background because of 
the source term in the right-hand side of the second equation 
in (\ref{BPSeq_MOmodel}). 
Thus Eqs.(\ref{eq_for_Sgm}) and (\ref{eq_for_phis}) are modified as 
\be
 \der_y\brkt{e^{6\sgm}\cN_I}+6k_I+\frac{\brkt{\frac{3}{2}k_I+m_I}J_0^2 
 \exp\brc{2(3k+2m)\cdot\int_0^y\dr y'\;\Re\,\vph_\Sgm}}
 {1-\frac{J_0^2}{4}\exp\brc{2(3k+2m)\cdot\int_0^y\dr y'\;\Re\,\vph_\Sgm}} 
 =0, \label{eq_for_Sgm2}
\ee
\be
 \vph^3 = \frac{J_0\exp\brc{(3k+2m)\cdot\int_0^y\dr y'\;\vph_\Sgm}}
 {4\sbk{1-\frac{J_0^2}{4}\exp\brc{2(3k+2m)\cdot\int_0^y\dr y'\;
 \Re\,\vph_\Sgm}}^{1/2}}, \;\;\;\;\;
 \vph^4 = k\cdot\Im\vph_\Sgm = 0. 
\ee
The arguments of $\cN$'s are $2\Re\,\vph_\Sgm$. 
The constant~$J_\pi$ is related to $J_0$ from the boundary condition of 
$\vph^3$ at $y=\pi R$ as 
\be
 J_\pi = J_0\exp\brc{(3k+2m)\cdot\int_0^{\pi R}\dr y\;\vph_\Sgm}. 
 \label{rel_Js}
\ee

In the case of $\nV=0$ (and $C_{000}=1$), 
the model is reduced to that of Ref.~\cite{MO}. 
If $J_0$ and $J_\pi$ are small enough, 
a BPS equation can be solved at the leading order of $J_0$-expansion as   
\bea
 \sgm \eql -ky+\cO(J_0^2), \nonumber\\
 \vph^3 \eql \frac{J_0}{4}\exp\brc{\brkt{\frac{3}{2}k+m}y}
 \brc{1+\cO(J_0^2)}, \nonumber\\
 \vph^4 \eql \Im\,\vph_\Sgm^0 =0, \label{BPSsol_MOmodel}
\eea
where $k\equiv k_0$ and $m\equiv m_0$. 
Here we have chosen the gauge where $\ge{y}{4}=1$.\footnote{
Thus $R$ is now the genuine radius of $S^1/Z_2$. } 
The relation~(\ref{rel_Js}) becomes 
\be
 J_\pi = J_0\exp\brc{\pi\brkt{\frac{3}{2}k+m}R}. \label{rel_Js2}
\ee
Due to this relation, the radius~$R$ is determined by the constants~$J_0$ 
and $J_\pi$, which implies that the radius is stabilized. 
Note that there is no arbitrary constant in the background solution. 
Namely no modulus exists in this model. 

Now we derive the 4D effective action by the procedure explained 
in the previous section. 
In the case of $\nVp\geq 1$, it is expressed from (\ref{4D_L_practical}) as 
\bea
 \cL^{\rm (4D)} \eql -3\pi R\int\dr^4\tht\;\abs{\phi}^2\left.\cN^{1/3}
 \right|_{y=0}\brkt{1-\abs{H}^2}^{2/3} \nonumber\\
 &&+\sbk{\int\dr^2\tht\;\phi^3\frac{1}{2}
 \brkt{J_0-J_\pi e^{-\pi\brkt{\frac{3}{2}k+m}T}}H+\hc}. 
\eea
Here note that $\cN^{1/3}|_{y=0}$ in this expression 
should be evaluated by using a solution of Eq.(\ref{EOM_Vp5}), 
which has a different form from the BPS equation~(\ref{eq_for_Sgm2}). 
We cannot obtain the correct effective action 
by using the BPS background solution in this model. 
 
In the case of $\nV=0$, we can perform the $y$-integral for 
(\ref{red_action1}) straightforwardly 
without a complicated procedure of integrating out $V^{I'}$~\cite{PST2} 
as mentioned in Sect.~\ref{nVp0_case}. 
In this model, (\ref{red_action1}) becomes 
\bea
 \cL^{\rm (4D)} \eql 
 -3\int\dr^4\tht\;\abs{\phi}^2\brc{\int_0^{\pi R}\dr y\;
 \brkt{-\der_y\tl{V}^0} e^{2k\tl{V}^0}
 \brkt{1-e^{-(3k-2m)\tl{V}^0}\abs{H}^2}^{2/3}} \nonumber\\
 &&+\sbk{\int\dr^2\tht\;\phi^3\frac{1}{2}\brc{J_0
 -e^{-\pi\brkt{\frac{3}{2}k+m}T}J_\pi}H+\hc}, 
\eea
where $T\equiv T^0$ is the radion multiplet. 
The $y$-integral can easily be performed and 
the following 4D Lagrangian is obtained. 
\be
 \cL^{\rm (4D)} = -3\int\dr^4\tht\;\abs{\phi}^2 \exp\brc{-K^{\rm (4D)}(T,H)/3} 
 +\sbk{\int\dr^2\tht\;\phi^3 P^{\rm (4D)}(T,H)+\hc}, 
\ee
where the K\"{a}hler potential~$K^{\rm (4D)}$ and 
the superpotential~$P^{\rm (4D)}$ are 
\bea
 K^{\rm (4D)}(T,H) 
 \eql -3\ln\brc{\frac{1-e^{-2\pi k\Re\,T}}{2k}
 -\frac{2\brkt{e^{\pi(k-2m)\Re\,T}-1}}{3(k-2m)}\abs{H}^2
 +\cO\brkt{\abs{H}^4}},  \nonumber\\
 P^{\rm (4D)}(T,H) \eql 
 \frac{1}{2}\brkt{J_0-J_\pi e^{-\pi\brkt{\frac{3}{2}k+m}T}}H. 
\eea
The scalar potential obtained from these functions indicates that 
there is a supersymmetric vacuum where both $t$ and $h$, 
which are the scalar components of $T$ and $H$, 
are stabilized to the values 
\be
 \vev{t} = \frac{\ln(J_\pi/J_0)}{\pi\brkt{\frac{3}{2}k+m}}, \;\;\;\;\;
 \vev{h} = 0. 
\ee
This is consistent with (\ref{BPSsol_MOmodel}) and (\ref{rel_Js2}).

\subsection{$Z_2$-even bulk mass for hypermultiplet} \label{sgm1_gauging}
So far we have considered only the case that the generators~$t_I$ act 
on each hypermultiplet~$(\Phi^{2\hat{a}-1},\Phi^{2\hat{a}})$ as $\sgm_3$. 
In such a case, the bulk masses for the hypermultiplets become $Z_2$-odd. 
Now we discuss a case that $t_I$ act on 
each hypermultiplet as $\sgm_1$.\footnote{
The $\sgm_2$-gauging leads to a similar result to 
that of the $\sgm_1$-gauging.}
We consider a simple case that $(\nV,\nC,\nH)=(0,1,1)$ and 
the physical hypermultiplet~$(\Phi^3,\Phi^4)$ has the following charge 
for $(V^0,\Sgm^0)$.  
\be
 igt_0 = \sgm_1\otimes\dgnl{0}{m}. 
\ee
This corresponds to an introduction of a $Z_2$-even mass~$m$ 
for the hypermultiplet in the flat 5D spacetime. 
The Lagrangian is written as 
\bea
 \cL \eql -3e^{2\sgm}\int\dr^4\tht\;\cV^0
 \brc{(\bar{\Phi}^1,\bar{\Phi}^2)\vct{\Phi^1}{\Phi^2}
 -(\bar{\Phi}^3,\bar{\Phi}^4)e^{-2mV^0\sgm_1}\vct{\Phi^3}{\Phi^4}}^{2/3} 
 \nonumber\\
 &&-2e^{3\sgm}\left[\int\dr^2\tht\;\left\{
 \Phi^1\brkt{\der_y+\frac{3}{2}\dot{\sgm}}\Phi^2
 -\Phi^3\brkt{\der_y+\frac{3}{2}\dot{\sgm}}\Phi^4 \right.\right. \nonumber\\
 &&\hspace{28mm}\left.\left.
 -m\Sgm^0\brkt{\brkt{\Phi^3}^2-\brkt{\Phi^4}^2}\right\}+\hc\right] \nonumber\\
 &&+e^{3\sgm}\sum_{\vth=0,\pi}
 \sbk{\int\dr^2\tht\;\brkt{\Phi^2}^2P^{(\vth)}\brkt{\frac{\Phi^4}{\Phi^2}}
 +\hc}\dlt(y-\vth R)+\cdots, 
 \label{sgm1_L}
\eea
where the ellipsis corresponds to the first line of (\ref{action2}), 
which is irrelevant to the following discussion. 
We have also introduced the boundary superpotentials. 

After gauging $\Sgm^0$ away, redefine the hypermultiplets 
as (\ref{fctout_phi}) and dropping the kinetic terms for 
the $Z_2$-odd multiplets, we obtain 
\be
 \der_y\phi=\der_y H=0 
\ee
from the equations of motion for $\phi^c$ and $H^c$. 
After the $y$-integration, we obtain 
the following K\"{a}hler potential~$K^{\rm (4D)}$ 
and superpotential~$P^{\rm (4D)}$.  
\bea
 K^{\rm (4D)}(T,H) \eql -3\ln\brc{-\int_0^{-\pi\Re\,T}\dr\tl{V}^0\;
 \brkt{1-\cosh(2m\tl{V}^0)\abs{H}^2}^{2/3}}, \nonumber\\ 
%
 P^{\rm (4D)}(T,H) \eql -\frac{\sinh(2\pi mT)}{2}H^2
 +\frac{1}{2}\brc{P^{(0)}(H)+P^{(\pi)}\brkt{\frac{H}{\cosh(\pi mT)}}}. 
 \label{Z2even_fcts}
\eea
We have taken the domain of the $y$-integration as $0\leq y <\pi R$. 
The first term in $P^{\rm (4D)}$ comes from a surface term at $y\to\pi R-$. 
It leads to a mass term for $H$ 
since $T$ is the radion multiplet and has a nonvanishing 
vacuum expectation value~$r$, which is the radius of the orbifold. 
Note that there is no counterpart to this mass term 
in the $\sgm_3$-gauging case. 
In the last term of $P^{\rm (4D)}$, 
we have used the relationship between $H_{\rm bulk}$ and $H|_{\pi R}$, 
\be
 H_{\rm bulk} = \cosh(\pi mT)H|_{\pi R}, 
\ee
which is obtained in a similar way to (\ref{rel_phis1}).

There is a consistency condition for the above 4D theory to be 
the effective theory of the original 5D theory. 
Namely $H$ and $T$ must be light enough comparing to the compactification 
scale~$m_{\rm KK}$, which is naturally thought to be lower than 
the 5D Planck scale~$M_5=1$. 
Thus the consistency condition is 
\be
 1>m_{\rm KK}\gg \frac{\sinh(2\pi mr)}{2}, 
\ee
which means that $\pi mr\ll 1$. 
Therefore we can expand the functions in (\ref{Z2even_fcts}) in terms of 
$\pi mT$. 
\bea
 K^{\rm (4D)}(T,H) \eql -3\ln\brc{\pi\Re\,T\brkt{1-\abs{H}^2}^{2/3}}
 +\cO\brkt{(\pi m\Re\,T)^2}, \nonumber\\
 P^{\rm (4D)}(T,H) \eql -\pi mTH^2+\frac{1}{2}\brc{
 P^{(0)}(H)+P^{(\pi)}(H)}+\cO\brkt{(\pi m\Re\,T)^2}. 
\eea
In fact, the equivalence between the case that the domain of 
the $y$-integral is taken as $0\leq y < \pi R$ 
and that of $0\leq y\leq\pi R$, which is mentioned below (\ref{def_L1}), 
holds only at the leading order of the above expansion. 
This is due to the existence of the mass term for $H$ in (\ref{Z2even_fcts}).

\subsection{$\nC=2$ case}
Here we consider the case that the compensator multiplet consists of 
two hypermultiplets, which is discussed in Sect.~\ref{nc2_case}. 
Eliminating $\Phi^1$ by using (\ref{EOM_SgmT}), 
the Lagrangian~(\ref{tc_action2}) becomes 
\bea
 \cL \eql -3e^{2\sgm}\int\dr^4\tht\;\brc{\cN(\cV)
 \brkt{\abs{\bdm{\Psi}^t\sgm_3\bdm{\Psi}^c}^2
 +\bdm{\Psi}^\dagger\sgm_3 e^{-2(\bt+\alp\tau)\cdot V}\bdm{\Psi}}
 \brkt{1+\bdm{\Psi}^{c\dagger}\sgm_3 e^{2(\bt+\alp\tau)\cdot V}\bdm{\Psi}^c}
 }^{1/3} \nonumber\\
 &&-e^{3\sgm}\sbk{\int\dr^2\tht\;\bdm{\Psi}^t\sgm_3
 \brc{\der_y+2(\bt+\alp\tau)\cdot\Sgm}\bdm{\Psi}^c+\hc}+\cdots, 
 \label{tc_action3}
\eea
where the ellipsis denotes terms corresponding to the first line of 
(\ref{action2}). 
\be
 \bdm{\Psi} = \vct{\Psi^3}{\Psi^5} \equiv 2\Phi^2\bdm{\Phi}, 
 \;\;\;\;\;
 \bdm{\Psi}^c = \vct{\Psi^4}{\Psi^6} \equiv 
 \frac{\bdm{\Phi}^c}{\Phi^2}. 
\ee

After gauging $\Sgm^I$ away by the transformation with $\Lmd_\Sgm^I$ 
in (\ref{gLmd1}), 
we redefine the hypermultiplets as follows. 
\bea
 \tl{\Psi}^3 \eql e^{-3\sgm}\phi^3, \;\;\;\;\;
 \tl{\Psi}^4 = \phi^c, \nonumber\\
 \tl{\Psi}^5 \eql e^{-3\sgm}\phi^3 H, \;\;\;\;\;
 \tl{\Psi}^6 = H^c. 
\eea
The new multiplets are defined so that $\phi$ has the Weyl weight one 
while the other multiplets~$(\phi^c,H,H^c)$ have zero weights.\footnote{
The Weyl weights of $\bdm{\Psi}$ and $\bdm{\Psi}^c$ are three and zero, 
respectively. } 
Then, dropping the kinetic terms for the $Z_2$-odd multiplets, 
the Lagrangian~(\ref{tc_action3}) becomes 
\be
 \cL = -3\int\dr^4\tht\;\abs{\phi}^2
 \cN^{1/3}(\cV)\brkt{\bdm{H}^\dagger\sgm_3e^{-2(\bt+\alp\tau)\cdot\tl{V}}
 \bdm{H}}^{1/3}
 -\sbk{\int\dr^2\tht\;\phi^3\bdm{H}\der_y\bdm{H}^c+\hc}, 
\ee
where 
\be
 \bdm{H} \equiv \vct{1}{H}, \;\;\;\;\;
 \bdm{H}^c \equiv \vct{\phi^c}{H^c}. 
\ee
From the equation of motion for $\bdm{H}^c$, we obtain 
\be
 \der_y\phi = \der_y H = 0. 
\ee
Thus $\phi$ and $H$ become 4D multiplets, which are the chiral compensator 
and physical multiplets, respectively. 
Therefore the 4D effective Lagrangian can be derived by integrating 
the following Lagrangian for $y$.  
\be
 \cL = -3\int\dr^4\tht\;\abs{\phi}^2\cN^{1/3}(\cV)
 e^{-\frac{2}{3}\bt\cdot\tl{V}}\brc{1-\abs{H}^2-2\alp\cdot\tl{V}
 \abs{1+H}^2}^{1/3}. 
\ee
We have used the nilpotency of $\tau$. 
In the case of $\nV=0$, the $y$-integral can easily be performed 
and the result in Ref.~\cite{PST2} is reproduced if $\bt_0=0$. 

Next we discuss the boundary actions. 
The chiral compensator and physical multiplets 
in the boundary Lagrangian~(\ref{nc2_Lbd}) are rewritten as~\footnote{
We should be careful not to confuse the powers with the upper indices. 
$\phi^3$ in (\ref{bd_fd_id}) is the cube of $\phi$ 
while the other numbers are values of the index~$a$. }
\bea
 \Phi^2\Phi^3 \eql \frac{\tl{\Psi}^3}{2} 
 = \frac{e^{-3\sgm}}{2}\phi^3, \nonumber\\
 Q \defa \frac{\Phi^5}{\Phi^3} = \frac{\tl{\Psi}^5}{\tl{\Psi}^3} 
 =H.  \label{bd_fd_id}
\eea
Again, the gauge-transformed quantities~$\tl{\Phi}^2\tl{\Phi}^3$ 
and $\tl{\Phi}^5/\tl{\Phi}^3$ are discontinuous at $y=\pi R$ 
because of the definition of $\Lmd_\Sgm^{I'}$ in (\ref{gLmd1}). 
\bea
 \lim_{y\to \pi R-}\brkt{\tl{\Phi}^2\tl{\Phi}^3} \eql 
 e^{-\pi\bt\cdot T}\brkt{\Phi^2\Phi^3}|_{\pi R}
 \brc{1-\pi\alp\cdot T-\pi\alp\cdot T\left.\frac{\Phi^5}{\Phi^3}
 \right|_{\pi R}}, \nonumber\\
 \lim_{y\to\pi R-}\frac{\tl{\Phi}^5}{\tl{\Phi}^3}
 \eql \frac{(1+\pi\alp\cdot T)(\Phi^5/\Phi^3)|_{\pi R}+\pi\alp\cdot T}
 {(1-\pi\alp\cdot T)-\pi\alp\cdot T(\Phi^5/\Phi^3)|_{\pi R}}. 
\eea
Thus the relation 
of the bulk values~$(\phi_{\rm bulk},H_{\rm bulk})$ and 
the boundary ones~$(\phi|_{\rm \pi R},H|_{\rm \pi R})$ can be read off as
\bea
 \phi^3|_{\pi R} \eql e^{\pi\bt\cdot T}\phi_{\rm bulk}^3
 \brc{1+\pi\alp\cdot T(1+H_{\rm bulk})}, \nonumber\\
 H|_{\pi R} \eql \frac{H_{\rm bulk}-\pi\alp\cdot T(1+H_{\rm bulk})}
 {1+\pi\alp\cdot T(1+H_{\rm bulk})}. 
 \label{nc2_bd_rel}
\eea
On the other hand, these chiral multiplets are continuous at $y=0$. 
We have to use these relations 
when the boundary actions~(\ref{nc2_Lbd}) exist. 
To make contact with the notation of Ref.~\cite{LOSW}, 
we sometimes redefine $H$ as~\cite{FKO} 
\be
 S \equiv \frac{1-H}{1+H}. 
\ee 
Then (\ref{nc2_bd_rel}) is rewritten as 
\bea
 \phi|_{\pi R} \eql e^{\frac{\pi}{3}\bt\cdot T}\phi_{\rm bulk}
 \brc{\frac{1+\brkt{S_{\rm bulk}+2\pi\alp\cdot T}}{1+S_{\rm bulk}}}^{1/3}, 
 \nonumber\\
 S|_{\pi R} \eql S_{\rm bulk}+2\pi\alp\cdot T. 
\eea

\section{Summary and comments} \label{comments}
We have discussed the dimensional reduction of 5D SUGRA on $S^1/Z_2$ 
keeping the $N=1$ off-shell structure. 
In such an off-shell dimensional reduction, the $Z_2$-odd multiplets 
play important roles. 
The key is the prescription~(\ref{prescription}). 
After this prescription, 
the equations of motion for the $Z_2$-odd multiplets provide 
constraints on the $Z_2$-even multiplets and extract the 4D zero modes 
from the latter. 

We would like to emphasize that the dimensional reduction procedure 
proposed in this paper is performed {\it in a background-independent way}. 
Namely we can derive the off-shell 4D action regardless of 
whether the background is supersymmetric or not. 
Thus we can apply our procedure to realistic brane-world models 
where the background breaks SUSY. 
Whether the background preserves SUSY or not can be judged 
by evaluating the vacuum in the off-shell 4D action. 
However there is a consistency condition for the derived 4D action 
to be the effective action of 5D SUGRA, that is, 
all the modes appearing in the 4D action must be light enough comparing 
to the compactification scale~$m_{\rm KK}$. 
We have to check this consistency condition after deriving the 4D action. 
Unlike the global SUSY case, the effective theory of 5D SUGRA 
is expressed as 4D SUGRA only up to $m_{\rm KK}$. 
Above $m_{\rm KK}$, there appear in the theory the Kaluza-Klein modes 
for the gravitational multiplet, which cannot be incorporated into 4D SUGRA. 
We would also like to emphasize that we have not used 
the superconformal gauge-fixing conditions in Appendix~\ref{SCgf} at all 
throughout the procedure. 
The gauge-fixing procedure is postponed until the dimensional reduction 
is completed. 

Another advantage of the off-shell dimensional reduction is that 
it can avoid the regularization problem of the orbifold singularity. 
As mentioned in Ref.~\cite{AS3}, we encounter the regularization-dependent 
quantities such as $\dlt^2(y)$ in the ordinary dimensional reduction, 
which is performed in the on-shell description. 
Such terms appear after eliminating the 5D auxiliary fields 
in the superconformal multiplets. 
In the off-shell dimensional reduction, on the other hand, 
the auxiliary fields are to be eliminated after the dimensional reduction. 
Thus the above singular terms do not appear when the auxiliary fields 
are eliminated because the orbifold singularity has already been integrated out. 
The only regularization dependent quantities encountered 
in the off-shell dimensional reduction 
are $\vep^2(y-\vth R)\dlt(y-\vth R)$ ($\vth=0,\pi$), 
which are assumed to be zero in this paper. 
The absence of the singular terms greatly simplify the calculations. 

Our work here corresponds to a reinterpretation and a modification of 
a method of Ref.~\cite{PST2}. 
So we will give some comments on the latter in the rest of this section.
In Ref.~\cite{PST2}, the 4D effective action is derived by the following 
procedure for $\nV'\geq 1$. 
Firstly, solve the BPS equations and find the BPS background solution. 
Then plug it into (\ref{red_action1}) and perform the $y$-integral. 
Here $\der_y\tl{V}^{I'}$ and $\tl{V}^{I'}$ in (\ref{red_action1}) 
are replaced with $\Re\,\Sgm^{I'}$ and $\int_0^y\dr y\;\Re\,\Sgm^{I'}$ 
since all the $Z_2$-odd multiplets have been dropped by hand. 
The backgrond solution contains real integration constants~$t_{\rm R}^{I'}$ 
for $\vph_\Sgm^{I'}$, which correspond to the moduli~$T^{I'}$. 
Thus the $y$-integrated action depends on $t_{\rm R}^{I'}$. 
Then $t_{\rm R}^{I'}$ are promoted to $\Re\,T^{I'}$, and 
the desired off-shell 4D action is obtained. 
In this procedure, the integration constants~$t_{\rm R}^{I'}$ are promoted 
to the superfields at the final step. 
However they can be promoted before plugging them into 
the action~(\ref{red_action1}) as we did in Ref.~\cite{AS2}. 
Due to the holomorphicity, 
the background solution~$\vph_\Sgm^{I'}=\vph_\Sgm^{I'}(t_{\rm R},y)$ 
is promoted as
\be
 \Sgm^{I'}=\vph_\Sgm^{I'}(T,y). 
\ee
In general, $\Re\,\Sgm^{I'}$ are not functions 
of only $\Re\,T^{I'}$ but also depend on $\Im\,T^{I'}$. 
Such $\Im\,T^{I'}$-dependence remains in the action after the $y$-integral. 
This contradicts the result obtained by promoting $t_{\rm R}^{I'}$ 
at the final step. 
This stems from the fact that $\tl{V}^{I'}$ have been dropped by hand. 
As shown in Sect.~\ref{VI_integrate}, 
the equations of motion for $\tl{V}^{I'}$ make 
the 4D K\"{a}hler potential~$K^{\rm (4D)}$ independent of $\Im\,T^{I'}$. 

The action~(\ref{red_action1}) has 
a shift symmetry~$\Sgm^{I'}\to\Sgm^{I'}+ic^{I'}$ 
(\ie, $T^{I'}\to T^{I'}+2iRc^{I'}$) for constants $c^{I'}$ 
if the boundary terms are absent. 
However this does not mean that $\Im\,\Sgm^{I'}$ are moduli. 
In fact, the BPS solution does not have such a symmetry 
as can be seen from $k\cdot\Im\vph_\Sgm=0$ in (\ref{eq_for_phis}). 
This is because the $\SUu$-gauge fixing condition fixes the phase of 
the compensator hyperscalar~$\vph^2$. 
From (\ref{fctout_phi}), $\Phi^2$ can be expressed as 
\be
 \Phi^2 = e^{-\frac{3}{2}\sgm}e^{3k\cdot\Lmd_\Sgm}\phi^{\frac{3}{2}}, 
\ee
and thus the constant shift of $\Im\,\Sgm^{I'}$ leads to 
a phase rotation of $\vph^2$, 
which is prohibited by the $\SUu$-gauge fixing. 
Namely the shift symmetry is a symmetry of the 5D conformal SUGRA but not 
of the 5D Poincar\'{e} SUGRA. 
In 4D conformal SUGRA, on the other hand, there is a symmetry of 
a phase rotation of the chiral compensator~$\phi$ 
until the 4D superconformal gauge fixing. 
Hence the shift symmetry is inherited as this phase rotation symmetry 
in the off-shell dimensional reduction. 
Therefore the discussion in Ref.~\cite{PST2} that 
$K^{\rm (4D)}$ becomes a function of only $\Re\,T^{I'}$ 
due to the shift symmetry is valid 
if the boundary terms are absent. 
In the presence of the boundary terms, however, 
such a discussion is no longer applicable because the $T$-dependence 
from them breaks the shift symmetry even before the gauge fixing. 
In that case, the $\Im\,T^{I'}$-independence of $K^{\rm (4D)}$ should be 
explained by the discussion in Sect.~\ref{VI_integrate}. 

Although the method of Ref.~\cite{PST2} uses the background information  
(\ie, the BPS solution), 
it is still valid as a practical method as mentioned in Sect.~\ref{practical}. 
However we have to note that what we should substitute 
into the action~(\ref{red_action1}) (or (\ref{4D_L_practical})) 
is a solution of (\ref{EOM_with_A0}) 
but not a BPS solution if the boundary terms exist.

\subsection*{Acknowledgement}
H.~A. and Y.~S. are supported by the Japan Society for the 
Promotion of Science for Young Scientists (No.0602496 and No.0509241).

\appendix

\section{Superconformal gauge-fixing conditions} \label{SCgf}
We collect the gauge-fixing conditions 
for the extraneous superconformal symmetries, 
\ie, the dilatation~$\bdm{D}$, $\SUu$, 
the conformal supersymmetry~$\bdm{S}$.\footnote{
The special conformal transformation~$\bdm{K}$ is already fixed 
in our $N=1$ description. }
We express them in terms of the components of the $N=1$ multiplets 
used in this paper. 

\bea
 V^I \eql \tht\sgm^{\udl{\mu}}\bar{\tht}W_\mu^I
 +i\tht^2\bar{\tht}\bar{\lmd}^I-i\bar{\tht}^2\tht\lmd^I
 +\frac{1}{2}\tht^2\bar{\tht}^2D^I, \nonumber\\
 \Sgm^I \eql \vph_\Sgm^I-\tht\chi_\Sgm^I-\tht^2\cF^I_\Sgm, \nonumber\\
 \Phi^a \eql \vph^a-\tht\chi^a-\tht^2\cF^a,
\eea
where $I=0,1,\cdots,\nV$ and $a=1,2,\cdots,2(\nC+\nH)$. 

The $\bdm{D}$ gauge is fixed by 
\be
 \cN(2\Re\,\vph_\Sgm) = \brkt{\ge{y}{4}}^3, \label{D_gauge1}
\ee
and 
\be
 \vph^a\dmx\bar{\vph}^b = 1, \label{D_gauge2}
\ee
where $\cN$ is the norm function defined by (\ref{def_normf}). 

The $\bdm{S}$ gauge is fixed by 
\be
 \cN_I(M)\lmd^I = \cN_I(M)\chi_\Sgm^I = 0, 
\ee
and 
\be
 \vph^a\dmx\bar{\chi}^b = \vph^a\dmx\rho_{bc}\chi^c = 0. 
\ee

\begin{description}
\item[$\bdm{\nC=1}$ case] \mbox{}\\
Combined with (\ref{D_gauge2}), the $\SUu$ gauge-fixing condition is written as
\be
 \vph^1 = 0, \;\;\;\;\;
 \vph^2 = \brkt{1+\sum_{a=3}^{2(\nH+1)}\abs{\vph^a}^2}^{1/2}. 
 \label{nc1_S_gauge}
\ee
Thus the compensator scalar~$\vph^2$ is expressed 
in terms of the physical hyperscalars~$\vph^a$ ($a\geq 3$). 

\item[$\bdm{\nC=2}$ case] \mbox{}\\
In this case, we cannot completely determine 
the compensators~$(\vph^1,\vph^2,\vph^3,\vph^4)$ 
in terms of the physical fields 
by using only the $\bdm{D}$ and $\SUu$ gauge-fixings. 
We need to use the gauge-fixing for $U(1)_T$. 
To see the situation, we focus on a case of $\nH=1$. 
We can obtain the following constraints on the hyperscalars~$\vph^a$ 
by picking up the lowest components of (\ref{EOM_VT}) and (\ref{EOM_SgmT}) 
after taking the Wess-Zumino gauge for $V_T$. 
\bea
 \abs{\vph^1}^2+\abs{\vph^3}^2-\abs{\vph^5}^2 
 \eql \abs{\vph^2}^2+\abs{\vph^4}^2-\abs{\vph^6}^2 = \frac{1}{2}, 
 \label{T_gauge1} \\
 \vph^1\vph^2+\vph^3\vph^4-\vph^5\vph^6 \eql 0. \label{T_gauge2}
\eea
In the second equality of (\ref{T_gauge1}), we have used 
the $\bdm{D}$ gauge-fixing condition~(\ref{D_gauge2}). 
Using the $\SUu$ gauge together with these constraints, 
All the compensator scalars are expressed in terms of the physical 
scalar fields~$(\phi_1,\phi_2)$ as follows~\cite{FKO}. 
\bea
 \vph^1 \eql 0, \;\;\;\;\;
 \vph^2 = \brkt{\frac{1-\abs{\phi_2}^2}
 {2(1-\abs{\phi_1}^2-\abs{\phi_2}^2)}}^{1/2}, \nonumber\\
 \vph^3 \eql \brkt{\frac{1}{2(1-\abs{\phi_2}^2)}}^{1/2}, \;\;\;\;\;
 \vph^4 = \vph^2\frac{\bar{\phi}_1\phi_2}{1-\abs{\phi_2}^2}, \nonumber\\
 \vph^5 \eql \vph^3\phi_2, \;\;\;\;\;
 \vph^6 = \vph^2\frac{\bar{\phi}_1}{1-\abs{\phi_2}^2}. \label{nc2_S_gauge}
\eea
Under the orbifold parity, $\phi_1$ and $\phi_2$ are odd and even, 
respectively. 

\end{description}

\section{Integrating out $\tl{V}^{I'}$: $3k_{I'}\neq 2m_{I'}$ case} 
\label{3kI_neq_2mI}
Here we will show how $\tl{V}^{I'}$ in (\ref{red_action1}) are integrated out 
in a case that $3k_{I'}\neq 2m_{I'}$ for some $I'$. 
Without loss of generality, we can label~$I'$ so that $3k_0\neq 2m_0$. 
In the case that $3k_{I'}=2m_{I'}$ for all $I'\neq 0$, 
we can apply the procedure explained in Sec.~\ref{VI_integrate} 
by taking the equations with $I'=1,2,\cdots,\nVp$ in (\ref{EOM_Vp4}) 
as $\nVp$ independent ones. 
Thus we assume that $3k_{I'}\neq 2m_{I'}$ for some value of $I'\neq 0$. 
Such values of $I'$ are denoted as $\Ip$. 
For $I'=\Ip$, we obtain the following equations from (\ref{EOM_Vp1}). 
\bea
 &&\hat{\cN}^{1/3}e^{2k\cdot\tl{V}}
 \brkt{1-e^{-(3k-2m)\cdot\tl{V}}\abs{H}^2}^{2/3} \nonumber\\
 \eql \frac{3c_{\Ip}}{2}
 \der_y\brc{\brkt{\frac{\hat{\cN}_{\Ip}}{3k_{\Ip}-2m_{\Ip}}
 -\frac{\hat{\cN}_0}{3k_0-2m_0}}
 \frac{e^{2k\cdot\tl{V}}}{\hat{\cN}^{2/3}}
 \brkt{1-e^{-(3k-2m)\cdot\tl{V}}\abs{H}^2}^{2/3}}, 
 \label{EOM_Vp_A}
\eea
where 
\be
 c_{\Ip} \equiv \brkt{\frac{3k_{\Ip}}{3k_{\Ip}-2m_{\Ip}}
 -\frac{3k_0}{3k_0-2m_0}}^{-1}.
\ee
If we define 
\be
 \cF_{\Ip} \equiv -\frac{9c_{\Ip}}{2\hat{\cN}^{2/3}}
 \brkt{\frac{\hat{\cN}_{\Ip}}{3k_{\Ip}-2m_{\Ip}}
 -\frac{\hat{\cN}_0}{3k_0-2m_0}}, \label{def_cF2}
\ee
it becomes a function of $v^i$ ($i=1,2,\cdots,\nVp$) defined by (\ref{def_vi}). 
Thus using (\ref{EOM_Vp_A}) and (\ref{def_cF2}) instead of 
(\ref{EOM_Vp3}) and (\ref{def_cF}), the second line of (\ref{4D_L1}) 
is replaced by 
\bea
 \cL^{\rm (4D)} \eql \int\dr^4\tht\;\abs{\phi}^2 \left\{
 \cF_{\Ip}(v_\pi)e^{-2\pi k_{J'}\Re\,T^{J'}}
 \brkt{1-e^{\pi(3k-2m)_{J'}\Re\,T^{J'}+2(\gh-\gc)_{I''}\tl{V}^{I''}}
 \abs{H}^2}^{2/3} \right.\nonumber\\
 &&\hspace{20mm}\left. 
 -\cF_{\Ip}(v_0)\brkt{1-e^{2(\gh-\gc)_{I''}\tl{V}^{I''}}\abs{H}^2}^{2/3}
 \right\}+\cdots, 
 \label{4D_L2}
\eea
where $v_0^i$ and $v_\pi^i$ are boundary values of $v^i$ 
at $y=0$ and $y=\pi R$, respectively. 

Now we will express $(v_0^i,v_\pi^i)$ in terms of the physical 4D modes. 
From (\ref{EOM_Vp2}), we also obtain 
\bea
 \frac{\der_y\brkt{\hat{\cN}_{\Ip}/\hat{\cN}^{2/3}}}{\brkt{(3k-2m)\cP}_{\Ip}}
 -\frac{\der_y\brkt{\hat{\cN}_0/\hat{\cN}^{2/3}}}{\brkt{(3k-2m)\cP}_0}
 \eql -\brc{\frac{6(k\cP)_{\Ip}}{\brkt{(3k-2m)\cP}_{\Ip}}
 -\frac{6(k\cP)_0}{\brkt{(3k-2m)\cP}_0}}\hat{\cN}^{1/3}, 
 \nonumber\\ \label{EOM_Vp_B}
\eea
where $(k\cP)_{I'}\equiv k_{J'}\cP^{J'}_{\;\;I'}$, and the arguments of 
$\hat{\cN}$'s and $\cP$ are $-\der_y\tl{V}$. 
Eq.(\ref{EOM_Vp_B}) is rewritten in the following form. 
\be
 -\der_y\tl{V}^{J'} = \sum_{i=1}^{\nVp}\cH_{\Ip J'i}(v^1,v^2,\cdots,v^{\nVp})
 \der_y v^i.  \label{der_yV2}
\ee
This is the same form as (\ref{der_yV1}). 
Thus we can repeat the procedure below (\ref{der_yV1}) 
by using (\ref{der_yV2}) for $I'=\Ip$. 
The only modification is to replace (\ref{FFs1}) with 
\bea
 &&\cF_1(v_\pi)e^{-2\pi k_{I'}\Re\,T^{I'}}
 \brkt{1-e^{\pi(3k-2m)_{I'}\Re\,T^{I'}+2(\gh-\gc)_{I''}\tl{V}^{I''}}
 \abs{H}^2}^{2/3} \nonumber\\
 &&-\cF_1(v_0)\brkt{1-e^{2(\gh-\gc)_{I''}\tl{V}^{I''}}
 \abs{H}^2}^{2/3} \nonumber\\
%
%
 \eql \cdots = \nonumber\\
 \eql \cF_{\nVp}(v_\pi)e^{-2\pi k_{I'}\Re\,T^{I'}}
 \brkt{1-e^{\pi(3k-2m)_{I'}\Re\,T^{I'}+2(\gh-\gc)_{I''}\tl{V}^{I''}}
 \abs{H}^2}^{2/3} \nonumber\\
 &&-\cF_{\nVp}(v_0)\brkt{1-e^{2(\gh-\gc)_{I''}\tl{V}^{I''}}
 \abs{H}^2}^{2/3}. 
\eea
Therefore, in this case, $(v_0^i,v_\pi^i)$ also depend on 
$H$ and $\tl{V}^{I''}$. 
\bea
 v_0^i \eql v_0^i(\Re\,T^0,\Re\,T^1,\cdots,\Re\,T^{\nVp},
 e^{2(\gh-\gc)_{I''}\tl{V}^{I''}}\abs{H}^2) \nonumber\\
 v_\pi^i \eql v_\pi(\Re\,T^0,\Re\,T^1,\cdots,\Re\,T^{\nVp},
 e^{2(\gh-\gc)_{I''}\tl{V}^{I''}}\abs{H}^2),  
\eea
where $i=1,2,\cdots,\nVp$. 
Plugging these into (\ref{4D_L2}), we obtain the off-shell 4D action.


\begin{thebibliography}{100}
 \bibitem{polchinski} J.~Polchinski, ``String Theory'', Vol.II, 
 Cambridge~Univ.~Press, Cambridege, UK, 1998.  
%
 \bibitem{maldacena} J.M.~Maldacena, {\it Adv.~Theor.~Math.~Phys.}~{\bf 2} 
 (1998) 231, {\it Int.~J.~Theor.~Phys.}~{\bf 38} (1999) 1113 
 [{\tt hep-th/9711200}]. 
%
 \bibitem{braneworld} E.A.~Mirabelli and M.E.~Peskin, \PR{D58} (1998) 065002 
 [{\tt hep-th/9712214}]; L.~Randall and R.~Sundrum, \NP{B557} (1999) 79 
 [{\tt hep-th/9810155}]; 
 G.F.~Giudice, M.A.~Luty, H.~Murayama and R.~Rattazzi, \JH{9812} (1998) 027 
 [{\tt hep-ph/9810442}]; D.E.~Kaplan, G.D.~Kribs and M.~Schmaltz, 
 \PR{D62} (2000) 035010 [{\tt hep-ph/9911293}]; 
 Z.~Chacko, M.~Luty, A.E.~Nelson and E.~Pont\'{o}n, 
 \JH{0001} (2000) 003 [{\tt hep-ph/9911323}]; 
 Z.~Chacko and M.A.~Luty, \JH{0105} (2001) 067 
 [{\tt hep-ph/0008103}]. 
%
 \bibitem{HW} P.~Ho\v{r}ava and E.~Witten, \NP{B460} (1996) 506 
 [{\tt hep-th/9510209}];{\bf B475} (1996) 94 [{\tt hep-th/9603142}]. 
%
 \bibitem{LOSW} A.~Lukas, B.A.~Ovrut, K.S.~Stelle and D.~Waldram, 
 \PR{D59} (1999) 086001 [{\tt hep-th/9803235}];
 \NP{B552} (1999) 246 [{\tt hep-th/9806051}]. 
%
 \bibitem{RS} L.~Randall and R.~Sundrum, \PRL{83} (1999) 3370 
 [{\tt hep-ph/9905221}]. 
%
 \bibitem{GP} T.~Gherghetta and A.~Pomarol, \NP{B586} (2000) 141 
 [{\tt hep-ph/0003129}]. 
%
 \bibitem{FLP} A.~Falkowski, Z.~Lalak and S.~Pokorski, \PL{B491} (2000) 172 
 [{\tt hep-th/0004093}]. 
%
 \bibitem{ABN} R.~Altendorfer, J.~Bagger and D.~Nemeschansky, 
 \PR{D63} (2001) 125025 [{\tt hep-th/0003117}]. 
%
 \bibitem{KU} M.~Kaku, P.K.~Townsend and P.~Van Nieuwenhuizen, \PRL{39} (1977)
 1109;  \PL{B69} (1977) 304; \PR{D17} (1978) 3179;
 T.~Kugo and S.~Uehara, \NP{B226} (1983) 49;\PTP{73} (1985) 235. 
%
 \bibitem{zucker} M.~Zucker, \NP{B570} (2000) 267 [{\tt hep-th/9907082}];
 \JH{0008} (2000) 016 [{\tt hep-th/9909144}].  
%
 \bibitem{KO2} T.~Kugo and K.~Ohashi, \PTP{105} (2001) 323 
 [{\tt hep-ph/0010288}]
%
 \bibitem{FKO} T.~Fujita, T.~Kugo and K.~Ohashi, \PTP{106} (2001) 671 
 [{\tt hep-th/0106051}]. 
%
 \bibitem{KO3} T.~Kugo and K.~Ohashi, \PTP{108} (2002) 203 
 [{\tt hep-th/0203276}]. 
%
 \bibitem{LS} M.A.~Luty and R.~Sundrum, \PR{D64} (2001) 065012 
 [{\tt hep-th/0012158}]. 
%
 \bibitem{PST1} F.~Paccetti Correia, M.G.~Schmidt and Z.~Tavartkiladze, 
 \NP{B709} (2005) 141 [{\tt hep-th/0408138}]. 
%
 \bibitem{AS1} H.~Abe and Y.~Sakamura, \JH{0410} (2004) 013 
 [{\tt hep-th/0408224}]. 
%
 \bibitem{LLP} W.D.~Linch III, M.A.~Luty and J.~Phillips, 
 \PR{D68} (2003) 025008 [{\tt hep-th/0209060}]. 
%
 \bibitem{BGGLLNP} I.L.~Buchbinder, S.J.~Gates,~Jr., H.S.~Goh, 
 W.D.~Linch III, M.A.~Luty, S.P.~Ng and J.~Phillips, 
 \PR{D70} (2004) 025008 [{\tt hep-th/0305169}]. 
%
 \bibitem{AS3} H.~Abe and Y.~Sakamura, \PR{D73} (2006) 125013 
 [{\tt hep-th/0511208}]. 
%
 \bibitem{PST2} F.~Paccetti Correia, M.G.~Schmidt and Z.~Tavartkiladze, 
 \NP{B751} (2006) 222 [{\tt hep-th/0602173}]. 
%
 \bibitem{HLW} L.J.~Hall, J.D.~Lykken and S.~Weinberg, 
 \PR{D27} (1983) 2359. 
%
  \bibitem{AS2} H.~Abe and Y.~Sakamura, \PR{D71} (2005) 105010 
 [{\tt hep-th/0501183}]. 
%
 \bibitem{WLP} B.~de Wit, P.G.~Lauwers and A.~Van Proeyen, 
 \NP{B255} (1985) 569. 
%
 \bibitem{BKV} E.~Bergshoeff, R.~Kallosh and A.~Van Proeyen, 
 \JH{0010} (2000) 033 [{\tt hep-th/0007044}]. 
%
 \bibitem{MP} D.~Mart\'{i} and A.~Pomarol, \PR{D64} (2001) 105025 
 [{\tt hep-th/0106256}]. 
%
 \bibitem{MO} N.~Maru and N.~Okada, \PR{D70} (2004) 025002 
 [{\tt hep-th/0312148}]. 
%
%
%
%
\end{thebibliography}
\end{document}